\documentclass[onecolumn,preprintnumbers,preprint,groupaddress,11pt]{revtex4}

\usepackage{amsmath}
\usepackage{graphics}
\usepackage{graphicx}
\usepackage[mathcal]{eucal}
\usepackage{color}
\usepackage{appendix}

\begin{document}
\preprint{UCI-TR-2010-10}

\title{Low Scale Non-universal, Non-anomalous $U(1)^{\prime}_{F}$ in a Minimal \\
Supersymmetric Standard Model} 

\author{Mu-Chun Chen}
\email[]{muchunc@uci.edu}
\author{Jinrui Huang}
\email[]{jinruih@uci.edu}
\affiliation{Department of Physics and Astronomy, University of California, Irvine, California 92697-4575, USA}

\date{\today}

\begin{abstract}
\label{abstract}
We propose a non-universal $U(1)^{\prime}_{F}$ symmetry combined with the Minimal Supersymmetric Standard Model. All anomaly cancellation conditions are satisfied without exotic fields other than three right-handed neutrinos. Because our model allows all three generations of chiral superfields to have different $U(1)^{\prime}_{F}$ charges, upon the breaking of the $U(1)^{\prime}_{F}$ symmetry at a low scale, realistic masses and mixing angles in both the quark and lepton sectors are obtained. In our model, neutrinos are predicted to be Dirac fermions and their mass ordering is of the inverted hierarchy type.  The $U(1)^{\prime}_{F}$ charges of the chiral super-fields also naturally suppress the $\mu$ term and automatically forbid baryon number and lepton number violating operators. 
While all flavor-changing neutral current constraints in the down quark and charged lepton sectors can be satisfied, we find that the constraint from $D^{0}-\overline{D}^{0}$ turns out to be much more stringent than the constraints from the precision electroweak data.    
\end{abstract}


\maketitle

\section{Introduction}
\label{sec:intro}

The $U(1)^{\prime}$ symmetry exists in many extensions of the Standard Model (SM). It can arise from a grand unified theory based on $SO(10)$ or $E_{6}$~\cite{ref:SO10, ref:E6, ref:E6Enrico}. In the presence of fermions or Higgs bosons with non-standard SM charges, a non-unifiable $U(1)^{\prime}$ symmetry can also exist~\cite{ref:nonUnifU1}. In addition, the $U(1)^{\prime}$ symmetries abound in many low energy effective theories of the string theories~\cite{ref:strM}. If the $U(1)^{\prime}$ breaking scale is at the TeV scale, the $Z^{\prime}$ gauge boson associated with the breaking of the $U(1)^{\prime}$ symmetry may be discovered at the early stages of the Large Hadron Collider (LHC) operation. 

Models with an extra $U(1)^{\prime}$ symmetry at the TeV scale are severely constrained by the flavor-changing-neutral-current (FCNC) processes and by the electroweak precision measurements. In the generation dependent $U(1)^{\prime}$ model, the off-diagonal terms in the Yukawa matrices can lead to FCNCs at the tree level through the exchange of the $Z^{\prime}$ gauge boson. The most stringent experimetal constraints are in the down-type quark sector from the measurements of $K^{0}-\overline{K}^{0}$ mixing and $B^{0}-\overline{B}^{0}$ mixing, and in the lepton sector from the non-observation of $\mu-e$ conversion, $\mu \rightarrow e^{+}e^{-}e^{+}$, $\tau \rightarrow e^{+}e^{-}e^{+}$ and $\tau \rightarrow \mu^{+}\mu^{-}\mu^{+}$. While recent measurement of $D^{0}-\overline{D}^{0}$ mixing places a bound on the $1-2$ mixing in the up-type quark sector, such limit is not as severe as those in the down-type quark sector. So far, there is no constraint on other mixing in the up-type quark sector.

To satisfy the FCNC constraints, most models~\cite{ref:PLRevZ', ref:Z'LHCRizzo, ref:Z'LHCRizzoInd} assume that the $U(1)^{\prime}$ charges are  universal across the three generations of Standard Model (SM) fermions. Due to the fact that the most stringent constraints from FCNCs appear in the processes that involve the first and second generations of fermions, a non-universal $U(1)^{\prime}$ at the TeV scale can be consistent with the experimental constraints on flavor violation, if the first two generations of the SM fermions have the same $U(1)^{\prime}$ charges and the flavor non-universality occurs between the third family charges and the charges of the first and second families of fermions~\cite{ref:FCNC}. 

In this paper, we relax the assumption of having universal charges for the first and second families. Instead,  all three generations of SM fermions are allowed to have different charges under the $U(1)^{\prime}$ symmetry. The FCNC constraints are satisfied by attributing the flavor mixing to the up-type quark and neutrino sectors, while having flavor diagonal down-type quark and charged lepton sectors, given that the down-type quark and charged lepton sectors are most stringently constrained. In this scenario, the $U(1)^{\prime}$ can play the role of a family symmetry \cite{ref:gauTrmNeuM, ref:SU(5)U(1), ref:highOpeWalter} that gives rise to realistic mass hierarchy and mixing angles among the SM fermions through the Froggatt-Nielsen (FN) mechanism~\cite{ref:frogNiel}. In addition, the $U(1)^{\prime}$ charge assignment naturally suppresses the $\mu$ term, and it forbids at the tree level baryon number and lepton number violating operators that could lead to proton decay.

This paper is organized as follows. In Section~\ref{sec:model} we present the flavor non-universal $U(1)^{\prime}$ model combined with MSSM. In particular we show how all gauge anomalies are cancelled and how realistic masses and mixing angles of all quarks and leptons (including the neutrinos) are generated. In addition, the implications for  the $\mu$ problem and proton decay is also discussed. Section~\ref{sec:ewpt} gives the parameter space of this model allowed by the most stringent experimental constraints. Phenomenological implications of our model for collider experiments are discussed in Section~\ref{sec:collider}. The mass spectrum of the super particles and the new phenomenology signatures in addition to the MSSM are discussed in section ~\ref{sec:susyMass}. Section~\ref{sec:conclude} concludes the paper. 

\section{The Model}
\label{sec:model}

In MSSM with three right-handed neutrinos, the superpotential for the Yukawa sector and Higgs sector that gives masses to all SM fermions and Higgs fields is given as follows,
\begin{equation}
\label{eqn:sPoten}
W = Y_uH_uQu^c + Y_dH_dQd^c + Y_eH_dLe^c + Y_{\nu}H_uL\nu^{c} + Y_{LL}LLH_uH_u + Y_{\nu\nu}\nu^c\nu^c + \mu H_uH_d + \mu^{\prime} \Phi \Phi^{\prime}\; .
\end{equation}
In the presence of an additional $U(1)^{\prime}_{F}$ symmetry under which various chiral super-fields are charged, the Yukawa matrices shown above are the effective Yukawa couplings generated through higher dimensional operators. As a result, they can be written as powers of the ratio of the flavon Higgs field, $\Phi$, that breaks the $U(1)^{\prime}_{F}$ symmetry, to the cutoff scale of the $U(1)^{\prime}_{F}$ symmetry, $\Lambda$,
\begin{equation}
\label{eqn:genYukawa1}
Y_{ij} \sim \biggl( y_{ij} \frac{\Phi}{\Lambda} \biggr)^{3|q_i+q_j+q_H|} \; .
\end{equation} 
Similarly, the effective $\mu$ term is generated by the higher dimensional operator and it is given by 
\begin{equation}
\label{eqn:genmu}
\mu \sim \biggl( \mu_{ud} \frac{\Phi}{\Lambda} \biggr)^{3|q_{H_u}+q_{H_d} - 1/3|} \Phi \; .
\end{equation} 
Here the chiral superfield $\Phi$ is a SM gauge singlet whose $U(1)^{\prime}_{F}$ charge is normalized to $-1/3$ in our model; $q_i$ and $q_j$  are the $U(1)^{\prime}_{F}$ charges of the chiral superfields of the $i$-th and $j$-th generations of quarks and leptons, while $q_H$ (which can be $q_{H_u}$ or $q_{H_d}$) denotes the $U(1)^{\prime}_{F}$ charges of the up- and down-type Higgses. Note that if $q_{i}+q_{j}+q_{H} < 0$ or $q_{H_{u}} + q_{H_{d}} < 1/3$, then instead of the $\Phi$ field, the field $\Phi^{\prime}$ whose $U(1)^{\prime}_{F}$ charge is $1/3$ is used respectively in Eq.(\ref{eqn:genYukawa1}) and (\ref{eqn:genmu}). The terms with non-integer $3|q_i+q_j+q_H|$ and $3|q_{H_u}+q_{H_d}|$ are not allowed in the superpotential given that the number of the flavon fields must be an integer. This thus naturally gives rise to texture-zeros in the Yukawa matrices. Once the scalar component $\phi$ ($\phi^{\prime}$) of the flavon superfield $\Phi$ ($\Phi^{\prime}$) acquires a vacuum expectation value (VEV), the $U(1)^{\prime}_{F}$ symmetry is broken. Upon the breaking of the $U(1)^{\prime}_{F}$ symmetry and the electroweak symmetry, the effective Yukawa couplings can be rewritten as,  
\begin{equation}
\label{eqn:genYukLam}
Y_{ij}^{eff} \sim \left(y_{ij} \epsilon \right)^{|q_i+q_j+q_H|},
\end{equation}
and the effective $\mu$ term is given by,
\begin{equation}
\label{eqn:genMuLam}
\mu \sim \left(\mu_{ud} \epsilon \right)^{|q_{H_u}+q_{H_d}-1/3|} <\phi> \; ,
\end{equation}
where $\epsilon \equiv \left( <\phi> / \Lambda \right)^3$ and $\epsilon^{\prime} \equiv \left(<\phi^{\prime}> / \Lambda \right)^3$. By choosing the expansion parameters $\epsilon$ and $\epsilon^{\prime}$ to be of the size of the Cabibbo angle $\sim 0.22$, we have found solutions to the charges that give rise to realistic fermion masses and mixing angles with all Yukawa couplings of order $y_{ij} \sim \mathcal{O}(1)$. One thing to address here is that although both $\epsilon$ and $\epsilon^{\prime}$ have to be of the size $\sim 0.22$, $<\phi>$ and $<\phi^{\prime}>$ are not necessarily to be the same due to the existence of the $\mathcal{O} (1)$ coefficients $y_{i,j}$ and $\mu_{ud}$. 
These charges also suppress the effective $\mu$ term by a factor of $\epsilon^{|q_{H_u}+q_{H_d}-1/3|}$. With $|q_{H_u}+q_{H_d}|$ having a value in the range of $\sim [1, 2]$, the effective $\mu$ term of the size of $100$ GeV can naturally arise. 

We then search for charges that satisfy all anomaly cancellation conditions and at the same time give rise to realistic masses and mixing angles for all SM fermions and a $\mu$ term of the right size. The details are discussed below.

\subsection{Anomaly Cancellation}

By expanding the gauge symmetry of MSSM with an additional $U(1)^{\prime}_{F}$ symmetry, there are six additional anomaly cancellation conditions. For all Higgs super-fileds in our model, we assume that they appear in conjugate pairs and therefore do not contribute to the gauge anomalies. As a result, only the charges of the three generations of matter fields are constrained by the anomaly cancellation conditions: \\
\begin{eqnarray}
\label{eqn:su3u1}
[SU(3)]^{2} U(1)^{\prime}_{F} & : &   \sum_{i} \left[ 2q_{Q_i} - (-q_{u_i}) - (-q_{d_i}) \right]  = 0 \; , 
\\
\label{eqn:su2u1}
[SU(2)_{L}]^{2} U(1)^{\prime}_{F} & : & \sum_{i} \left[ q_{L_i} + 3q_{Q_i}\right] = 0  \; , 
\\
\left[U(1)_{Y}\right]^{2} U(1)^{\prime}_{F} & : & 
\sum_{i} \biggl[ 2 \times 3 \times \biggl( \frac{1}{6} \biggr)^2 q_{Q_i} - 3 \times \biggl( \frac{2}{3} \biggr)^2 (-q_{u_i} ) - 3 \times \biggl( -\frac{1}{3}\biggr)^2 (-q_{d_i}) \label{eqn:u1y2u1}  \\
& & \qquad \qquad  + 2 \times \biggl(-\frac{1}{2}\biggr)^2 q_{L_i} - (-1)^2 (-q_{e_i}) \biggr] = 0 \; , \nonumber 
\\
\left[U(1)_{F}^{\prime}\right]^{2} U(1)_{Y} & : & 
\displaystyle \sum_{i} \biggl[ 2 \times 3 \times \biggl( \frac{1}{6} \biggr) q_{Q_i}^2 - 3 \times \biggl( \frac{2}{3} \biggr) \times (-q_{u_i})^2 - 3 \times \biggl(-\frac{1}{3} \biggr) (-q_{d_i})^2 \label{eqn:u1yu12}\\
& & \qquad \qquad + 2 \times \biggl(-\frac{1}{2}\biggr)(q_{L_i})^2 - (-1)(-q_{e_i})^2 \biggr] = 0 \; , 
\nonumber \\
\label{eqn:u1grav}
U(1)^{\prime}_{F}-\mbox{gravity} & : & 
\displaystyle \sum_{i} \left[ 6q_{Q_i} + 3q_{u_i} + 3q_{d_i} + 2q_{L_i} + q_{e_i} + q_{N_i}\right] = 0 \; , 
\\
\label{eqn:u13}
[U(1)^{\prime}_{F}]^{3} & : &  \hspace{-0.05in}
\sum_{i} \left[ 3 \bigl( 2 (q_{Q_i})^3 - (-q_{u_i})^3 - (-q_{d_i})^3\bigr) + 2(q_{L_i})^3 - (-q_{e_i})^3 - (-q_{N_i})^3\right] = 0\;. 
\end{eqnarray}
where $q_{Q_{i}}$, $q_{u_{i}}$, $q_{d_{i}}$, $q_{L_{i}}$, $q_{e_{i}}$, and $q_{N_{i}}$ denote, respectively, the charges of the quark doublet, iso-singlet up-type quark, iso-singlet down-type quark, lepton doublet, iso-singlet charged lepton, and right-handed neutrino of the $i$-th generation. 
To further reduce the number of parameters, we also assume that the fields $Q_{i}$, $u_{i}^c$, and $e_{i}^c$ have the same $U(1)^{\prime}$ charges, $q_{Q_i} = q_{u_i} = q_{e_i} \equiv q_{t_i}$, and that the fields $L_{i}$ and $d_{i}^c$ have the same $U(1)^{\prime}$ charges, $q_{L_i} = q_{d_i} \equiv q_{f_i}$, as motivated by the $SU(5)$ unification~\cite{ref:SU(5)U(1)}. With these assignments, the above six anomaly cancellation conditions reduce to the following three independent ones, 
\begin{eqnarray} 
\label{eqn:anomaly1} 
\frac{1}{2} \displaystyle \sum_{i} q_{f_i} + \frac{3}{2} \displaystyle \sum_{i} q_{t_i} & = & 0 \; , \\
\label{eqn:anomaly2} 
5 \displaystyle \sum_{i} q_{f_i} + 10 \displaystyle \sum_{i} q_{t_i} + \displaystyle \sum_{i} q_{N_i} & = & 0 \; , \\
\label{eqn:anomaly3}
5 \displaystyle \sum_{i} q_{f_i}^3 + 10 \displaystyle \sum_{i} q_{t_i}^3 + \displaystyle \sum_{i} q_{N_i}^3 & = & 0 \; . 
\end{eqnarray}
The first two anomaly conditions, Eqs.(\ref{eqn:anomaly1}) and (\ref{eqn:anomaly2}), are satisfied automatically with the following parametrization of the $U(1)^{\prime}_{F}$ charges, 
\begin{eqnarray} 
\begin{array}{lll} q_{t_1} & = & -\frac{1}{3} q_{f_1} - 2a \; , \nonumber \\
q_{t_2} & = & -\frac{1}{3} q_{f_2} + a + a^{\prime} \; ,  \nonumber \\
q_{t_3} & = & -\frac{1}{3} q_{f_3} + a - a^{\prime} \; ,  \end{array} \nonumber \,\,\qquad
\begin{array}{lll} q_{N_1} & = & -\frac{5}{3} q_{f_1} - 2b \; , \nonumber  \\
q_{N_2} & = & -\frac{5}{3} q_{f_2} + b + b^{\prime} \; , \nonumber \\
q_{N_3} & = & -\frac{5}{3} q_{f_3} + b - b^{\prime} \; . \end{array}
\label{eqn:param}
\end{eqnarray}
where parameters $a$, $a^{\prime}$, $b$, and $b^{\prime}$ characterize the charge splittings between different generations of $q_{t_i}$ and $q_{N_i}$. The charges $q_{f_i}$ and charge splitting parameters, $a$, $a^{\prime}$, $b$, and $b^{\prime}$, are  determined by the cubic equation Eq.(\ref{eqn:anomaly3}) and the observed fermion masses and mixings as shown in the following section.

\subsection{Fermion Masses and Mixings}
\label{sec:massMix}

The $U(1)^{\prime}_{F}$ charges give the following up-type quark Yukawa matrix,
\begin{eqnarray}
\label{eqn:upYukawa}
Y_U & \sim & \left(\begin{array}{lll} (\epsilon)^{|2q_{t_1}+q_{H_u}|} & (\epsilon)^{|q_{t_1}+q_{t_2}+q_{H_u}|} & (\epsilon)^{|q_{t_1}+q_{t_3}+q_{H_u}|} \\ (\epsilon)^{|q_{t_1}+q_{t_2}+q_{H_u}|} & (\epsilon)^{|2q_{t_2}+q_{H_u}|} & (\epsilon)^{|q_{t_2}+q_{t_3}+q_{H_u}|} \\ (\epsilon)^{|q_{t_1}+q_{t_3}+q_{H_u}|} & (\epsilon)^{|q_{t_2}+q_{t_3}+q_{H_u}|} & (\epsilon)^{|2q_{t_3}+q_{H_u}|} \end{array} \right)  \; ,
\end{eqnarray}
and the Yukawa matrix of the down-type quarks is given by, 
\begin{eqnarray}
\label{eqn:downYukawa}
Y_D & \sim & \left(\begin{array}{ccc} (\epsilon)^{|q_{t_1}+q_{f_1}+q_{H_d}|} & (\epsilon)^{|q_{t_1}+q_{f_2}+q_{H_d}|} & (\epsilon)^{|q_{t_1}+q_{f_3}+q_{H_d}|} \\ (\epsilon)^{|q_{t_2}+q_{f_1}+q_{H_d}|} & (\epsilon)^{|q_{t_2}+q_{f_2}+q_{H_d}|} & (\epsilon)^{|q_{t_2}+q_{f_3}+q_{H_d}|} \\ (\epsilon)^{|q_{t_3}+q_{f_1}+q_{H_d}|} & (\epsilon)^{|q_{t_3}+q_{f_2}+q_{H_d}|} & (\epsilon)^{|q_{t_3}+q_{f_3}+q_{H_d}|} \end{array} \right)  \; .
\end{eqnarray}
(It is again to be understood that if the arguments of the absolute values are negative, then $\epsilon^{\prime}$ should be utilized instead of $\epsilon$.) Because the top quark is heavy, we assume that its mass term is generated at the renormalizable level, and thus
\begin{equation}
\label{eq:q1}
2q_{t_3}+q_{H_u} = 0 \; .
\end{equation}
To avoid the tree level FCNCs while allowing all three generations of chiral super-fields to have different $U(1)^{\prime}_{F}$ charges, we attribute all flavor mixings in the up-type quark and neutrino sectors with down-type quark and charged lepton sectors being flavor diagonal. Ideally the texture-zeros in the down-type quark and charged lepton sectors are generated due to the non-integer exponents as determined by the $U(1)^{\prime}_{F}$ charges. Nevertheless, no solution is found that can give diagonal down-type quark and charged lepton sectors and at the same time satisfy all other constraints in the model.  We therefore impose an additional $Z_{8}$ symmetry to forbid the off diagonal elements in the down-type quark and charged lepton mass matrices. The transformation properties of various chiral superfields are summarized in  Table ~\ref{tbl:z8Charg}.
\begin{table}[tbh!]
\begin{tabular}{c|c|c|c|c|c|c|c|c|c|c|c|c} \hline \hline 
Field & $d_1^c, e_1^c$ & $d_2^c, e_2^c$ & $d_3^c, e_3^c$ & $Q_1, u_1^c$ & $Q_2, u_2^c$ & $Q_3, u_3^c$ & $\nu_1^c$ & $\nu_2^c$ & $\nu_3^c$ & $H_u, H_d$ & $\Phi$ & $\Phi^{\prime}$ \\ \hline
$Z_{8}$ Parity & $-1$ & $i$ & $e^{-\frac{\pi i}{4}}$ & $1$ & $1$ & $1$ & $-1$ & $1$ & $e^{-\frac{\pi i}{4}}$ & $1$ & $e^{\frac{\pi i}{4}}$ & $1$ \\ \hline \hline 
\end{tabular}
\caption{The $Z_{8}$ parity of the chiral superfields. $d_i^c$ is the down-type quark singlet chiral superfield with family index $i$, $e_i^c$ is the charged lepton singlet, $Q_i$ is the quark doublet chiral superfield,   $u_i^c$ is the up-type quark singlet, $L_i$ is the lepton doublet, and $\nu_i^c$ is the neutrino singlet. 
}  
\label{tbl:z8Charg}
\end{table}

Realistic mass and mixing spectrum is found with 
\begin{eqnarray}
\label{eq:q2}
q_{f_1} - q_{f_3} = 22/3 \; , \qquad & q_{f_2} - q_{f_3} = 11/3 \; , \\ 
 q_{H_u}+q_{H_d} = -5/3 \; \qquad & a^{\prime} = -7/18 \; , \nonumber
 \end{eqnarray}
and also 
\begin{equation}
\label{eq:q3}
q_{t_3} + q_{f_3} + q_{H_d} = 1/3 \; , 
\end{equation}
which naturally accounts for the mass hierarchy between the top quark and the bottom quark, leading to a prediction of $\tan \beta = v_u/v_d \sim 25$, with
\begin{equation}
\left< H_{u} \right> = v_{u} \; , \quad \left< H_{d} \right> = v_{d} \; ,
\end{equation}
being the VEVs of the neutral scalar components of the Higgs supermultiplets.  All elements in the up-type quark Yukawa matrix has $Z_8$ parity of $+1$, and thus they are all allowed. The elements in the down-type quark Yukawa matrix have the following transformation properties under the $Z_8$ parity, 
\begin{eqnarray}
\label{eqn:downYukawaParity}
P_D & \sim & \left(\begin{array}{lll} 1 & e^{\frac{3\pi}{4}i} & e^{-\frac{\pi}{4}i} \\ e^{\frac{3\pi}{4}i} & 1 & e^{-\frac{\pi}{4}i} \\ e^{\frac{3\pi}{2}i} & e^{\frac{3\pi}{2}i} & 1 \end{array} \right) \; ,
\end{eqnarray}
As a result, only the diagonal elements in the down-type quark Yukawa matrix are allowed. The resulting effective Yukawa matrices of the up- and down-type quarks are thus given, in terms of $\epsilon$ or $\epsilon^{\prime}$, as
\begin{eqnarray}
\label{eqn:yu}
Y_U & \sim & \left(\begin{array}{lll} (\epsilon^{\prime})^{22/3} & (\epsilon^{\prime})^{17/3} & (\epsilon^{\prime})^{11/3} \\ (\epsilon^{\prime})^{17/3} & (\epsilon^{\prime})^{4} & (\epsilon^{\prime})^{2} \\ (\epsilon^{\prime})^{11/3} & (\epsilon^{\prime})^{2} & 1 \end{array} \right) \; , \\
\label{eqn:yd}
Y_D & \sim & 
\left(\begin{array}{lll} (\epsilon)^{4} & 0 & 0 \\ 
0 & (\epsilon)^{2} & 0 \\ 0 & 0 & (\epsilon)^{1/3} \end{array} \right) \; .
\end{eqnarray}
Due to the $SU(5)$-inspired charge assignment, we also have $Y_E = Y_{D}^{T}$. Eqs.(\ref{eqn:yu}) and (\ref{eqn:yd}) give rise to realistic masses of up- and down-type quarks and charged leptons as well as all CKM matrix elements. Since no mixing appears in the down-type  quark and the charge lepton sectors, all FCNC constraints are satisfied.

With the above charge assignment, the $[U(1)^{\prime}_{F}]^{3}$ anomaly cancellation condition is satisfied if  
\begin{equation}
\label{eq:q4}
q_{f_3} = \frac{-10240 - 63525b - 13365b^2 - 486b^3 + 9075b^{\prime} - 2970bb^{\prime} - 1485b^{\prime 2} + 486bb^{\prime 2}}{10(304+1485b+243b^2 - 495b^{\prime} + 81b^{\prime 2})} \; , 
\end{equation} 
for any $b$ and $b^{\prime}$. The values of $b$ and $b^{\prime}$ are determined by the neutrino sector. 
The Dirac, left-handed, and right-handed Majorana Neutrino mass terms are given, respectively, as follows,
\begin{eqnarray}
\label{eqn:neutDiracYukawa}
Y_N & \sim & \left(\begin{array}{lll} (\epsilon)^{|q_{f_1}+q_{N_1}+q_{H_u}|} & (\epsilon)^{|q_{f_1}+q_{N_2}+q_{H_u}|} & (\epsilon)^{|q_{f_1}+q_{N_3}+q_{H_u}|} \\ (\epsilon)^{|q_{f_2}+q_{N_1}+q_{H_u}|} & (\epsilon)^{|q_{f_2}+q_{N_2}+q_{H_u}|} & (\epsilon)^{|q_{f_2}+q_{N_3}+q_{H_u}|} \\ (\epsilon)^{|q_{f_3}+q_{N_1}+q_{H_u}|} & (\epsilon)^{|q_{f_3}+q_{N_2}+q_{H_u}|} & (\epsilon)^{|q_{f_3}+q_{N_3}+q_{H_u}|} \end{array} \right) \; ,
\\
\label{eqn:neutMajLeYukawa}
Y_{LLHH} & \sim & \left(\begin{array}{lll} (\epsilon)^{|2q_{f_1}+2q_{H_u}|} & (\epsilon)^{|q_{f_1}+q_{f_2}+2q_{H_u}|} & (\epsilon)^{|q_{f_1}+q_{f_3}+2q_{H_u}|} \\ (\epsilon)^{|q_{f_2}+q_{f_1}+2q_{H_u}|} & (\epsilon)^{|2q_{f_2}+2q_{H_u}|} & (\epsilon)^{|q_{f_2}+q_{f_3}+2q_{H_u}|} \\ (\epsilon)^{|q_{f_3}+q_{f_1}+2q_{H_u}|} & (\epsilon)^{|q_{f_3}+q_{f_2}+2q_{H_u}|} & (\epsilon)^{|2q_{f_3}+2q_{H_u}|} \end{array} \right) 
\; , \\
\label{eqn:neutMajRiYukawa}
Y_{NN} & \sim & \left(\begin{array}{lll} (\epsilon)^{|2q_{N_1}|} & (\epsilon)^{|q_{N_1}+q_{N_2}|} & (\epsilon)^{|q_{N_1}+q_{N_3}|} \\ (\epsilon)^{|q_{N_2}+q_{N_1}|} & (\epsilon)^{|2q_{N_2}|} & (\epsilon)^{|q_{N_2}+q_{N_3}|} \\ (\epsilon)^{|q_{N_3}+q_{N_1}|} & (\epsilon)^{|q_{N_3}+q_{N_2}|} & (\epsilon)^{|2q_{N_3}|} \end{array} \right) \; .
\end{eqnarray}
By choosing 
\begin{equation}
\label{eq:q5}
b = 55/8 \; , \qquad b^{\prime} = -347/18 \; ,
\end{equation} 
only the Dirac neutrino mass matrix is allowed since all elements in the left-handed and right-handed neutrino Majorana mass matrices are non-integers. Furthermore, elements of the Dirac neutrino mass matrix have the following transformation properties under the $Z_{8}$ parity, 
\begin{eqnarray}
\label{eqn:yvParity}
P_N & \sim & \left(\begin{array}{lll} -1 & 1 & 1 \\ -1 & 1 & e^{\frac{\pi}{4}i} \\ -1 & 1 & i \end{array} \right) \; .
\end{eqnarray}
Therefore, only the second column and the (3,1) element are allowed, leading to the following Dirac neutrino mass matrix,
\begin{equation}
\label{eqn:yv}
Y_N \sim  \left(\begin{array}{lll} 0 & (\epsilon^{\prime})^{49/3} & (\epsilon)^{85/3} \\ 0 & (\epsilon^{\prime})^{20} & 0 \\ 0 & (\epsilon^{\prime})^{71/3} & 0 \end{array} \right) \; .
\end{equation}
With the following $Y_{ij}$ coefficients of the order of unity, we obtain
\begin{eqnarray}
\label{eqn:yvPrime}
Y_{N} & = & \left(\begin{array}{lll} \; 0 \; & \;(0.8526)^{49}(\epsilon^{\prime})^{21} \; & \; (1.186633)^{85}(\epsilon)^{55/3} \; \\ \; 0 \; & \; (1.02678)^{60}(\epsilon^{\prime})^{19} \; & \; 0 \; \\ \; 0 \; & \; (1.105762e^{\frac{i\pi}{71}})^{71}(\epsilon^{\prime})^{19} \; & \; 0 \; \end{array} \right) \; .
\end{eqnarray}
The matrix $Y_{N} Y_{N}^{\dagger}$ is given by, 
\begin{eqnarray}
\label{eqn:yvPrime}
Y_{N} Y_{N}^{\dagger} & = & \left(\begin{array}{lll} (\epsilon)^{110/3} & (\epsilon^{\prime})^{40} & -(\epsilon)^{40} \\ (\epsilon^{\prime})^{40} & (\epsilon^{\prime})^{38} & -(\epsilon^{\prime})^{38} \\ -(\epsilon^{\prime})^{40} & -(\epsilon^{\prime})^{38} & (\epsilon^{\prime})^{38} \end{array} \right) \; .
\end{eqnarray}
The resulting neutrino mixing pattern arising from the matrix $Y_{N}$ given above is close to the tri-bimaximal mixing pattern. The three absolute masses are predicted to be, 
\begin{equation}
m_{\nu_1} \simeq 0.048214 \; \mbox{eV} \; ,\quad  
m_{\nu_2} \simeq 0.048988 \; \mbox{eV} \; ,\quad
m_{\nu_3} \simeq 0 \; .
\end{equation}
These three masses give the following values for the squared mass differences
\begin{equation}
|\Delta m_{atm}^{2}| = 2.40 \times 10^{-3} eV^{2} \; , \quad \Delta m_{\odot}^{2} =7.52 \times 10^{-5} eV^2\; ,
\end{equation}
which  satisfy the neutrino experimental results ~\cite{ref:neutinoPhysics, ref:neutrinoTheta13} and predict the inverted mass ordering for the light neutrinos. 

The $U(1)^{\prime}_{F}$ charges that correspond to the parameters given in Eqs.(\ref{eq:q1}), (\ref{eq:q2}), (\ref{eq:q3}), (\ref{eq:q4}), and (\ref{eq:q5}) are summarized in Table~\ref{tbl:u1Charge}.
\begin{table}[t!]
\begin{tabular}{c|c|c|c|c|c|c|c} \hline \hline
Field & $d_1^c, L_1$ & $d_2^c, L_2$ & $d_3^c, L_3$ & $Q_1, u_1^c, e_1^c$ & $Q_2, u_2^c, e_2^c$ & $Q_3, u_3^c, e_3^c$ & 
\\ \hline
$U(1)_{F}^{\prime}$ charge 
& $q_{f_{1}} = \frac{102857}{15585}$ 
& $q_{f_{2}} = \frac{45712}{15585}$ 
& $q_{f_{3}} = -\frac{3811}{5195}$ 
& $q_{t_{1}} = -\frac{42944}{15585}$ 
& $q_{t_{2}} = -\frac{16969}{15585}$ 
& $q_{t_{3}} = \frac{14201}{15585}$ 
& 
\\
\hline \hline
Field &  $\nu_1^c$ & $\nu_2^c$ & $\nu_3^c$ & $H_u$ & $H_d$ & $\Phi$ & $\Phi^{\prime}$ \\ \hline
$U(1)_{F}^{\prime}$ charge 
& $q_{N_{1}} = -\frac{17778}{1039}$ 
& $q_{N_{2}} = -\frac{21934}{1039}$ 
& $q_{N_{3}} = \frac{73424}{3117}$ 
& $q_{H_{u}} = -\frac{28402}{15585}$ 
& $q_{H_{d}} = \frac{2427}{15585}$ 
& $q_{\Phi} = -\frac{1}{3}$ 
& $q_{\Phi^{\prime}} = \frac{1}{3}$ \\ \hline \hline
\end{tabular}
\caption{The $U(1)^{\prime}_{F}$ charges of the chiral superfields.}  
\label{tbl:u1Charge}
\end{table}

\subsection{Implications for the $\mu$ problem and Proton Decay}
\label{sec:mu}

Due to the presence of the $U(1)_{F}^{\prime}$ symmetry, the $\mu$ parameter in our model is given by, 
\begin{equation}
\label{eqn:muMass}
|\mu|^2 = \frac{1}{2} \biggl[\frac{M_{H_u}^2 - M_{H_d}^2 + (q_{H_u} - q_{H_d})C}{\cos 2 \beta} 
- \biggl(M_{z}^2 + M_{H_u}^2 + M_{H_d}^2 + (q_{H_u} + q_{H_d})C \biggr) \biggr] \,\, .
\end{equation}
The parameter $C$ is defined as
\begin{equation}
\label{eqn:varA}
C = g_{z'}^2 (q_{H_u} v^2 \sin^2 \beta + q_{H_d} v^2 \cos^2 \beta - q_{\Phi} u^{2} \cos 2\psi) \; , 
\end{equation}
where $v^{2} = v_{u}^{2} + v_{d}^{2}$ while $u$ and $\psi$ are defined through $\left< \phi \right> = u \sin\psi $ and $\left< \phi^{\prime} \right> = u \cos \psi$,  
with $\phi$ and $\phi^{\prime}$ being the scalar component of the chiral superfield $\Phi$ and $\Phi^{\prime}$, respectively. The more detailed description of the Higgs sector of our model is given in Section~\ref{sec:higgs-sector}. 

Generally, a delicate cancellation between the Higgs masses and a $\mu$ term of the weak scale is required in order to obtain the observed $M_Z$~\cite{ref:muParaDawson}. This is known as the $\mu$ problem~\cite{ref:muProblem}. 
In our $U(1)^{\prime}_{F}$ model, the charge assignment of $H_{u}$ and $H_{d}$ naturally suppresses the $\mu$ term by a factor of $\sim (\epsilon)^{4/3} \sim 0.133$ with respective to $\left< \phi \right> \sim$ TeV scale. This thus naturally gives a $\mu$ term of the order of $\sim \mathcal{O}(100)$ GeV  while having $\mu_{ud} \sim \mathcal{O}(1)$.

The $U(1)^{\prime}_{F}$ charge assignment of the chiral superfields also automatically forbids lepton number and baryon number violating operators at the tree level. In general, lepton number and baryon number violating operators 
\begin{eqnarray}
\label{eqn:lepVi}
W_{\Delta L = 1} & = & \frac{1}{2} \lambda^{ijk} L_i L_j e_k^c + \lambda^{\prime ijk} L_i Q_j d_k^c + \mu^{\prime i}L_i H_u \; , \\
\label{eqn:baryVi}
W_{\Delta_B = 1} & = & \frac{1}{2} \lambda^{\prime \prime ijk} u_i^c d_j^c d_k^c \; ,
\end{eqnarray}
are allowed by supersymmetry, and they can lead to proton decay processes, {\it e.g.} 
\begin{equation}
p \rightarrow e^{+} \pi^0, \; e^{+} K^0, \; \mu^{+} \pi^0, \; \mu^{+} K^0, \; \nu \pi^0, \; \mbox{or} \; \;  \nu K^0 \; .
\end{equation}
To avoid those operators, the usual way is to impose the conservation of the R-parity, which is defined as $P_{R} = (-1)^{3(B-L)+2s}$ with $B$ being the baryon number, $L$ being the lepton number, and $s$ being the spin of the particle ~\cite{ref:susyPrimer}. Without imposing the R-parity, in our model the $U(1)^{\prime}_{F}$ charges automatically forbid these operators since these operators are not $U(1)^{\prime}_{F}$ gauge invariant.

\section{Experimental Constraints}   
\label{sec:ewpt}
\subsection{Electroweak Precision Constraints}

\begin{figure}[t!]
\includegraphics[scale=0.8, angle = 0, width = 80mm, height = 50mm]{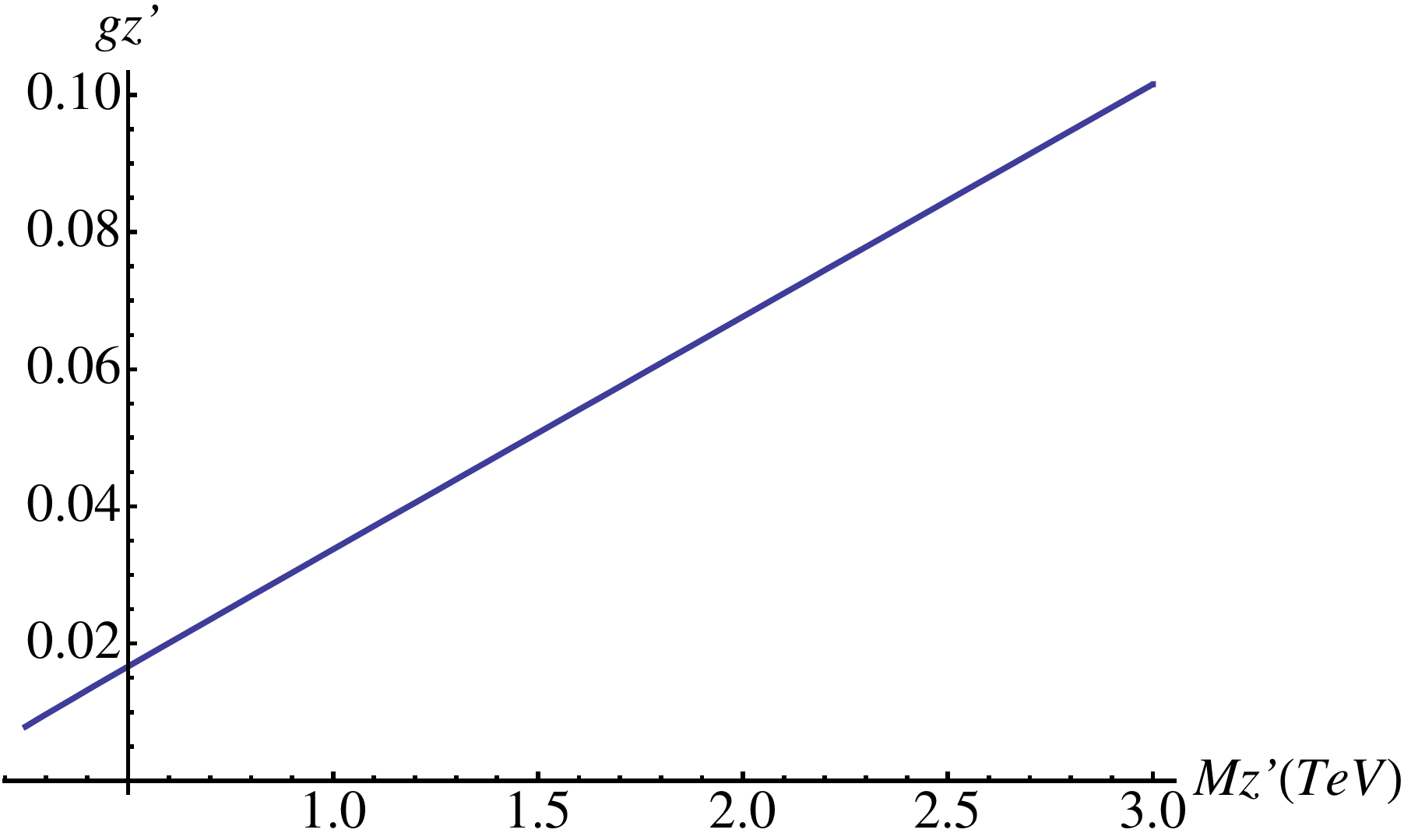}
	\caption{The maximally allowed value of the $U(1)^{\prime}_{F}$ gauge coupling, $g_{z'}$, as a function of the mass of the $Z^{\prime}$ gauge boson, $M_{Z^{\prime}}$, derived from the constraints on the $\rho$ parameter.}
	\label{fig:couplingMass}
\end{figure}

Because $H_{u}$ and $H_{d}$ both are charged under $U(1)_{F}^{\prime}$, tree-level kinetic mixing between the $Z$ and $Z^{\prime}$ gauge bosons exists in our model. The $Z-Z^{\prime}$ mixing contributes to the $\rho$ parameter and therefore it is very severely constrained ~\cite{ref:ewCons, ref:mixZZ, ref:CDFz'Con1}. The kinetic terms of the Higgses lead to mass terms for the gauge bosons, $Z_{F}^{\prime}$, $B_{Y}$, and $W^{3}$, which are associated with $U(1)^{\prime}_{F}$, $U(1)_{Y}$, and the $T^{3}$ generator of the $SU(2)_{L}$ symmetry, respectively,  
\begin{eqnarray}
\frac{1}{4} \biggl[ v_u^2 (-g_2 W^{3} + g_1 B_{Y} + 2 q_{H_u} g_{z'} Z_{F}^{\prime})^2 + v_d^2 (g_2 W^3 - g_1 B_{Y} + 2 q_{H_d} g_{z'} Z_{F}^{\prime})^2 \biggr]  \; , \nonumber
\end{eqnarray}
and $g_2$, $g_1$, and $g_{z'}$ are the $SU(2)_{L}$,  $U(1)_Y$, and  $U(1)^{\prime}_{F}$ gauge coupling constants. After diagonalizing the mass matrix for the gauge bosons, we obtain three physical eigenstates which are identified as the photon, the $Z$ boson, and the $Z^{\prime}$ boson
\begin{eqnarray}
\label{eqn:massBoson}
A & = & \frac{1}{\sqrt{g_2^2 + g_1^{2}}}(g_1 W^3 + g_2 B_{Y}) \; , \\
Z^{SM} & = & \frac{1}{\sqrt{g_2^2 + g_1^{2}}}(g_2 W^3 - g_1 B_{Y}) \; , \\
Z & = & Z^{SM} + \delta_{ZZ^{\prime}} Z_{F}^{\prime} \; , \\
Z^{\prime} & = & Z_{F}^{\prime}  - \delta_{ZZ^{\prime}} Z^{SM} \; .
\end{eqnarray}
The masses of the physical gauge bosons are given by
\begin{eqnarray}
M_{Z} & = & \sqrt{\frac{g_2^2 + g_1^{2}}{2}} \sqrt{v_u^2 + v_d^2}(1 + O(\delta_{ZZ^{\prime}}^2)) \; , \\
M_{Z^{\prime}} & = & g_{z'} \sqrt{2(q_{H_u}^2 v_u^2 + q_{H_d}^2 v_d^2 + 2 q_{\Phi}^2 u_{\phi}^2)}(1 + O(\delta_{ZZ^{\prime}}^2)) \; , 
\end{eqnarray}
and the term 
\begin{equation}
\delta_{ZZ^{\prime}} = \frac{\Delta M_{ZZ^{\prime}}^2}{M_{Z^{\prime}}^2 - M_{Z}^2} \; , \quad \mbox{with} \quad 
\Delta M_{ZZ^{\prime}}^2  =  g_{z'} \sqrt{g_2^2 + g_1^{2}}(q_{H_d} v_d^2 - q_{H_u} v_u^2) \; ,
\end{equation}
is the Higgs induced kinetic mixing between $Z$ and $Z^{\prime}$, which is severely constrained by the precision electroweak data. 
The electroweak precision measurements indicate that the $\rho$ parameter is very close to $1$ and the experimentally allowed deviation, $\Delta \rho$, is  
given by, 
\begin{equation}
|\Delta \rho| = \biggl| \frac{(\Delta M_{ZZ^{\prime}}^2)^2}{(M_{Z^{\prime}}^2 - M_{Z}^2)M_{Z}^2} \biggr| < 0.00023 \; .
\end{equation}
In the $U(1)^{\prime}_{F}$ model, $\Delta \rho$ is given by, 
\begin{equation}
\Delta \rho = \frac{4 g_{z'}^2 (q_{H_d} - q_{H_u} \tan^2 \beta)^2}{\biggl(\frac{M_{Z^{\prime}}^2}{M_{Z}^2} - 1\biggr)(g_2^2 + g_1^{2})(1 + \tan^2 \beta)^2} \; .
\end{equation}
The experimental limit on $\Delta \rho$ then is translated into the following constraints on the $U(1)^{\prime}$ gauge coupling $g_{z'}$ and the $Z^{\prime}$ mass with $\tan \beta = 25$ in our model,
\begin{equation}
g_{z'} < \sqrt{\frac{0.00023}{24.213637}\biggl( \frac{M_{Z^{\prime}}^2}{M_{Z}^2} - 1 \biggr)} \; .
\end{equation}
As shown in Figure~\ref{fig:couplingMass}, for a relatively light $Z^{\prime}$ mass of 600 GeV, the gauge coupling $g_{z'}$ must be smaller than $\lesssim 0.02$ in order to satisfy the precision electroweak constraint on the $\rho$ parameter. With increasing $Z^{\prime}$ mass, the maximally allowed value for $g_{z'}$ increases, to a very good approximation, linearly. 

We note that, in addition to the Higgs-induced contribution discussed above, the $Z-Z^{\prime}$ mixing can also be generated by an explicit kinetic mixing term in the Lagrangian and by the renormalization group evolution ~\cite{ref:rgeLoop}. There thus exists possible cancellation among these contributions to the $\rho$ parameter, allowing the constraints on $g_{z'}$ and $M_{Z^{\prime}}$ to be loosened. 

\subsection{Constraints from Flavor-Changing Neutral Currents}

Following the formalism in ~\cite{ref:FCNC}, the neutral current Lagrangian in the gauge eigenstates can be written as
\begin{equation}
  \mathcal{L}_{NC}=-eJ^{\mu}_{\mbox{\tiny em}}A_{\mu} - g_2 J^{\mu} Z_{\mu}^{SM}
           - g_{z'} J^{\prime \; \mu} Z_{F \; \mu}^{\prime}\;,
\end{equation}
where $e = g_1 g_2/\sqrt{g_1^2 + g_2^2}$ and $\cos \theta_w = g_2/\sqrt{g_1^2 + g_2^2} \,\, , \sin \theta_w = g_1/\sqrt{g_1^2 + g_2^2}$ and $\theta_w$ is the weak mixing angle in SM. The currents are defined as
\begin{eqnarray}
  J^{\mu}&=&\sum_i \bar{\psi}_i \gamma_{\mu} 
    \left[ \epsilon_{i}^{\psi_L} P_L + \epsilon_{i}^{\psi_R} P_R\right]\psi_i\;,\label{J1}\\[1ex]
  J^{\prime \; \mu}&=&\sum\limits_{i,j} \bar{\psi}_i \gamma_{\mu} 
    \left[ \epsilon_{ij}^{\prime \; \psi_L} P_L 
     + \epsilon_{ij}^{\prime \; \psi_R} P_R\right]\psi_j\;,\label{J2}
\end{eqnarray}
where the summations are taken over all quarks and leptons, $\psi_{i,j}$, and $P_{R,L}=\frac{1}{2} (1\pm\gamma_5)$ are the projection operators. The gauge coupling constants of the SM Z boson are given by 
\begin{equation}
  \epsilon_{i}^{\psi_R}=-\sin^2\theta_wQ_{e_{i}} \;, \qquad
  \epsilon_{i}^{\psi_L}=t_3^i-\sin^2\theta_w Q_{e_{i}} \;,
\end{equation}
where $t_3^i$ and $Q_{e_{i}}$ are the third component of the weak isospin and the electric charge of fermion $i$, respectively. The gauge coupling constants to $Z^{\prime}$ are denoted by $\epsilon_{ij}^{\prime \; \psi_{L,R}}$. Flavor-changing neutral currents immediately arise if the matrices $\epsilon^{\prime \; \psi_{L,R}}$ are non-diagonal. FCNCs can also be induced by the quark and lepton mixing, if $\epsilon^{\prime \; \psi_{L,R}}$ are diagonal but have non-universal elements. The fermion Yukawa matrices $Y_{\psi}$ in the weak eigenstate basis can be diagonalized by unitary matrices $V^{\psi}_{R,L}$
\begin{equation}
  Y_{\psi,diag}=V^{\psi}_R\,Y_{\psi}\,{V_L^{\psi}}^{\dagger} \; .
\end{equation}
Hence, the $Z^{\prime}$ coupling matrices in the fermion mass eigenstate basis are
\begin{equation}
  B^{\psi_L}\equiv
    \left(V_L^{\psi} \epsilon^{\prime \; \psi_L} {V_L^{\psi}}^{\dagger}\right)\;,
    \qquad \qquad
  B^{\psi_R}\equiv
    \left(V_R^{\psi} \epsilon^{\prime \; \psi_R} {V_R^{\psi}}^{\dagger}\right)\;.
  \label{Bij}
\end{equation}
The currents $J^{\mu}$ and $J^{\prime}$ in the mass eigenstates of the fermions can then be written as
\begin{eqnarray}
  J_{m}^{\mu}&=&\sum_i \bar{\psi}_{L_i}^{m} \gamma_{\mu} \epsilon_i^{\psi_L} \psi_{L_i}^{m} + \bar{\psi}_{R_i}^{m} \gamma_{\mu} \epsilon_i^{\psi_R} \psi_{R_i}^{m} \;,\label{J1}\\[1ex]
  J_{m}^{\prime \; \mu}&=&\sum\limits_{i,j} \bar{\psi}_{L_i}^{m} \gamma_{\mu} 
    B_{ij}^{\psi_L} \psi_{L_j}^{m} + \bar{\psi}_{R_i} \gamma_{\mu} B_{ij}^{\psi_R} \psi_{R_j}^{m}\; , \label{J2}
\end{eqnarray}
and the neutral current interaction in the bases of the mass eigenstates of the fermions, $Z$ and $Z^{\prime}$ as
\begin{equation}
  \mathcal{L}_{NC}^{m}= - g_1 \left[\cos\theta J_{m}^{\mu} 
      + \frac{g_{z'}}{g_1}\sin\theta J_{m}^{\prime \; \mu}\right] Z_{\mu}
    - g_1\left[\frac{g_{z'}}{g_1}\cos\theta J_{m}^{\prime \; \mu} 
      - \sin\theta J_{m}^{\mu}\right] Z_{\mu}^{\prime}\;, 
      \label{LZ}
\end{equation}
with $\theta$ being the $Z-Z^{\prime}$ mixing angle and $\sin \theta \sim \delta_{ZZ^{\prime}}$. 
The unitary matrices $V^{\psi}_{L}$ are constrained by the CKM matrix in the left-handed quark sector through the relation
\begin{equation}
  V_{CKM} = 
    V_L^u {V_L^d}^{\dagger}\;,
  \label{BCKM}
\end{equation}
and equivalently for the lepton sector by the PMNS matrix, 
\begin{equation}
V_{PMNS} = {V_{L}^{\nu}}^{\dagger} V_{L}^{e} \; .
\end{equation}

The flavor-changing neutral current interactions are severely constrained by various experiments ~\cite{ref:flavorPhysics} such as rare meson decays and neutral meson mixings, in particular  $D^0-\bar{D}^0$, $K^0-\bar{K}^0$, and $B^0-\bar{B}^0$ mixing. Generally, the mass splitting $\Delta m_P$ between neutral mesons $P^0$ and $\bar{P}^0$ through interaction shown in Eq.(\ref{LZ}) can be approximately written as (neglecting the negligible $Z-Z^{\prime}$ mixing effect)
\begin{equation}
\Delta m_P=\left(\frac{g_{z'}}{M_{Z^{\prime}}}\right)^2 m_P F_P^2\left\{
  \frac{1}{3}\mbox{Re}\left[\left(B_{ij}^{q_L}\right)^2
    +\left(B_{ij}^{q_R}\right)^2\right]-
  \left[\frac{1}{2}+
    \frac{1}{3}\left(\frac{m_P}{m_{q_i}+m_{q_j}}\right)^2\right]
  \mbox{Re}\left(B_{ij}^{q_L}B_{ij}^{q_R}\right)\right\}\;,
\end{equation}
where $m_P$ and $F_P$ are the mass and decay constant of the meson, respectively.
Since the mass eigenstates and the gauge eigenstates of the down-type quarks and the charged lepton sector are the same, there are no FCNCs in the down-type quark and charged-lepton sectors through the $Z^{\prime}$ exchange at the tree level. For the up-type quark sector, the only available experimental constraints is for the first two families and the most stringent one is from $D^0-\bar{D}^0$ mixing. In our model, the $B^{\psi_L}$ matrice is
\begin{equation}
  B^{u_L}\equiv
    \left(V_{CKM} \epsilon^{\prime \; u_L} V_{CKM}^{\dagger}\right)\;, \\
    \qquad 
  \label{Bij}
\end{equation}
and $B^{u_R} = - B^{u_L}$ since $u_L$ and $u_R$ carry opposite $U(1)^{\prime}_F$ charges. Choosing the standard parametrization of the CKM matrix, we have 
\begin{eqnarray}
V_{CKM} = \left(\begin{array}{ccc} C_{12}C_{13} & S_{12}C_{13} & S_{13}e^{-i\delta_{13}} \\ -S_{12}C_{23} - C_{12}S_{23}S_{13}e^{i\delta_{13}} & C_{12}C_{23} - S_{12}S_{23}S_{13}e^{i\delta_{13}} & S_{23}C_{13} \\ S_{12}S_{23} - C_{12}C_{23}S_{13}e^{i\delta_{13}} & -C_{12}S_{23} - S_{12}C_{23}S_{13}e^{i\delta_{13}} & C_{23}C_{13} \end{array}\right)  \; ,
\end{eqnarray}
where $C_{ij} \equiv \cos \theta_{ij}$, $S_{ij} \equiv \sin \theta_{ij}$, and $\theta_{ij}$ are mixing angles. As a result of the $U(1)^{\prime}$ symmetry, our model can accommodate the experimentally observed values, $\theta_{12} = 13.04^{\circ}$, $\theta_{13} = 0.201^{\circ}$, and $\theta_{23} = 2.38^{\circ}$. The CP phase in the CKM matrix is not determined in our model. In the following we assume $\delta_{13} =68.75^{\circ}$.
The matrix $\epsilon^{\prime \; u_L} $  is then given by
\begin{eqnarray}
\epsilon^{\prime \; u_L} = \left(\begin{array}{ccc} -42944/15585 & 0 & 0 \\ 0 & -16969/15585 & 0 \\ 0 & 0 & 14201/15585 \end{array}\right) \; , 
\end{eqnarray}
and the B matrix is then
\begin{eqnarray}
\label{matrx:BuL}
B^{u_L} = \left(\begin{array}{ccc} -2.6705 & 0.366226 - 0.000486337i & -0.590674 - 0.0117013i \\ 0.266226 + 0.000486337i & -1.1701 & 0.220285 + 0.00127662i \\ -0.590674 + 0.0117013i & 0.220285 - 0.00127662i & 0.769274 \end{array}\right) \; ,
\end{eqnarray}
which leads to the following contribution to the mass splitting due to the $Z^{\prime}$ exchange, 
\begin{eqnarray}
\Delta m_D & = & \left(\frac{g_{z'}}{M_{Z^{\prime}}}\right)^2 m_D F_P^2\left\{
  \frac{1}{3}\mbox{Re}\left[\left(B_{12}^{u_L}\right)^2
    +\left(B_{12}^{u_R}\right)^2\right]-
  \left[\frac{1}{2}+
    \frac{1}{3}\left(\frac{m_D}{m_{u}+m_{c}}\right)^2\right]
  \mbox{Re}\left(B_{12}^{u_L}B_{12}^{u_R}\right)\right\} \nonumber \\
  & = & 42.9483 \; g_{z'}^2 \left(\frac{1 \; \mbox{TeV}}{M_{Z^{\prime}}}\right)^2 \; \mbox{eV} \; .
\end{eqnarray}
The experimental constraint on the mass spliting of $D^0-\bar{D}^0$ system is $|\Delta m_D| \leq 1.56 \times 10^{-5}$ eV ~\cite{ref:pdgDDbar}. Therefore, we obtain conservatively the constraint 
\begin{equation}
\frac{g_{z'}}{M_{Z^{\prime}} \, (\mbox{in TeV})} < 6.027 \times 10^{-4} \; , 
\end{equation}
which is much more stringent than the constraints from the precision electroweak data. Hence, taking $g_{z'} = 0.02$, the mass of the $Z^{\prime}$ has to satisfy $M_{Z^{\prime}} > 33.18$ TeV. Similar to the $Z-Z^{\prime}$ mixing, there exist additional contributions from the sparticle sector to the $D$ meson mixing which can potentially cancel the contribution due to the $Z^{\prime}$ exchange~\cite{ref:FCNCSUSY}, allowing the constraints on the $g_{z'}$ and $M_{Z^{\prime}}$ to be loosened. In addition, the contributions due to the long range effects in the standard model, which are intrinsically non-perturbative and thus have large uncertainty, potentially also can relax the constraints, when included.   

We also note that in our model the contributions to the D meson leptonic and semi-leptonic decay modes through $Z^{\prime}$ mediation are negligible.

\section{Collider Signatures}
\label{sec:collider}

\begin{figure}[t!]
\includegraphics[scale=0.8, angle = 90, width = 80mm, height = 50mm]{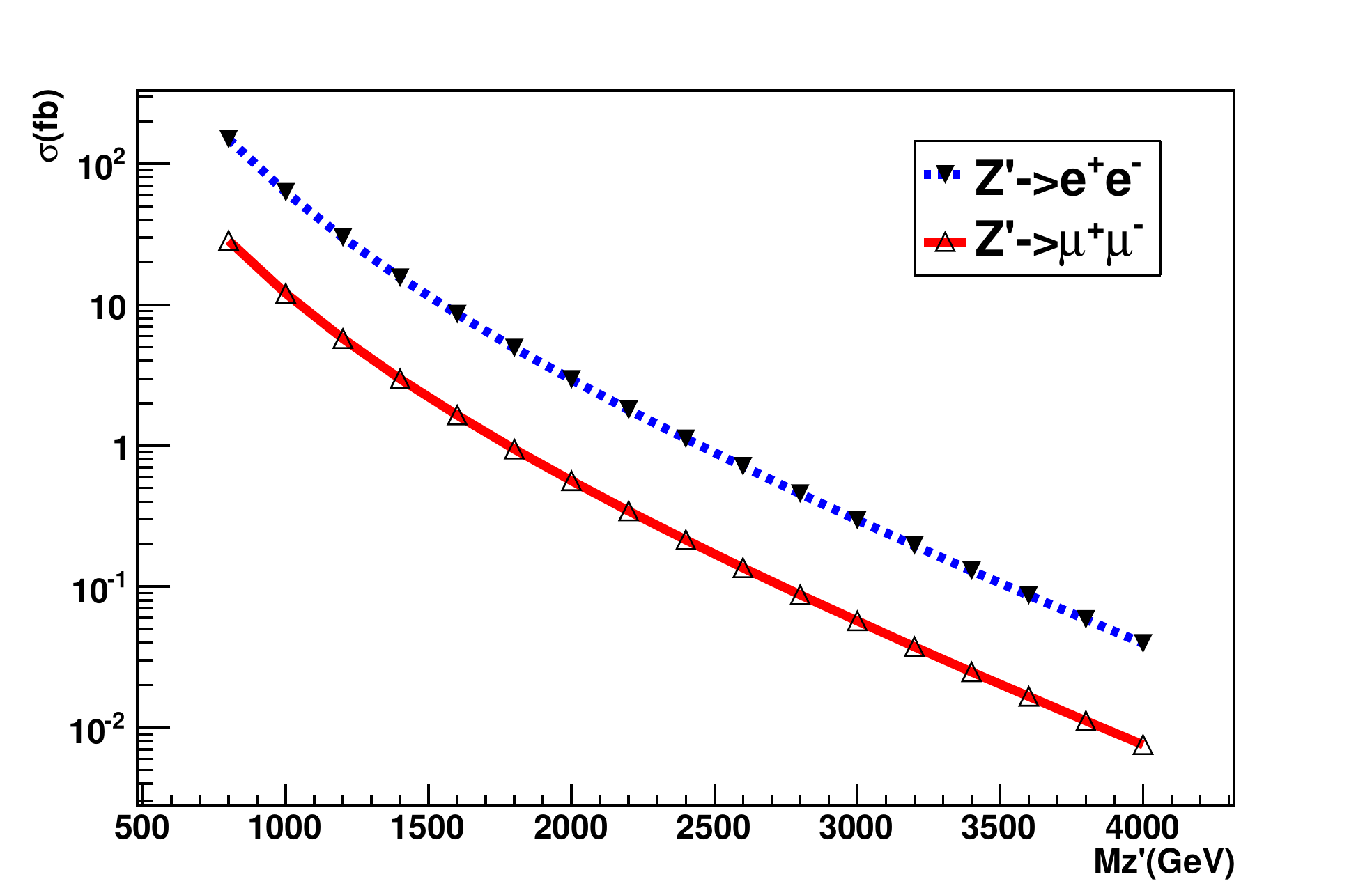}
	\caption{Cross Sections of $Z^{\prime} \rightarrow e^{+}e^{-}$ and $Z^{\prime} \rightarrow \mu^{+}\mu^{-}$ for different values of the $Z^{\prime}$ mass at the LHC with the center of mass energy $\sqrt{s} = 14$ TeV and $g_{z'} = 0.02$.}
	\label{fig:xSectDiLeptonLHC}
\end{figure}

The $Z^{\prime}$ gauge boson associated with the $U(1)^{\prime}_{F}$ breaking has a mass on the order of a TeV and therefore it can be produced at the collider experiments. A recent updated limit specific for our model from CDF $Z^{\prime}$ searches at CDF  can be found in Ref.~\cite{ref:CDFlimit}.  
With the assumption that the contribution to the $D^0-\bar{D}^0$ mixing from the $Z^{\prime}$ exchange can be compensated from other new physics such as SUSY, we can relax the severe bound from the $D^0-\bar{D}^0$ mixing, allowing a $M_{Z^{\prime}}$ of 1 TeV to be viable. Even though the $U(1)^{\prime}_{F}$ gauge coupling constant is constrained to be small in order to satisfy the precision electroweak constraints, due to the large $U(1)^{\prime}_{F}$ charges of the three generations of chiral super-fields, the $Z^{\prime}$ may still be discovered through the resonances of the di-muon or di-electron decay channels. Fig.~\ref{fig:xSectDiLeptonLHC} shows the cross sections of $Z^{\prime} \rightarrow e^{+}e^{-}$ and $Z^{\prime} \rightarrow \mu^{+}\mu^{-}$ at the LHC as a function of the $Z^{\prime}$ mass with the center of mass energy $\sqrt{s} = 14$ TeV and $g_{z'} = 0.02$.  

Due to the fact that the $U(1)^{\prime}$ charges of the superfields are generation dependent, in addition to the flavor conserving $Z^{\prime}$ decay channels, some flavor violating decay channels are also allowed in our model. These are discussed in the following.

\subsection{Flavor Conserving Decays}

Since the three generations of superfields have non-universal $U(1)^{\prime}$ charges, the branching fractions among different $Z^{\prime}$ leptonic decay channels are very distinguishable due to the large charge splittings. The $Z^{\prime}$ partial decay width for different channels at the tree level is given by,
\begin{equation}
\Gamma(Z^{\prime} \rightarrow \psi_i \bar{\psi}_j) = \frac{N_c M_{Z^{\prime}}}{24\pi}\left(|-g_{z'} \cos \theta B_{ij}^{\psi_L} + g_1 \sin \theta \delta_{ij} \epsilon_{i}^{\psi_L}|^2 + |-g_{z'} \cos \theta B_{ij}^{\psi_R} + g_1 \sin \theta \delta_{ij} \epsilon_{i}^{\psi_R}|^2\right),
\label{zDeWid}
\end{equation}
in which the color factor $N_{c}$ is 1 for leptons and 3 for quarks.
The flavor conserving processes contributing to the $Z^{\prime}$ decay width through $Z-Z^{\prime}$ mixing is relatively small. Therefore, the $Z^{\prime}$ decay width can be estimated by 
\begin{equation}
\Gamma(Z^{\prime} \rightarrow \psi_i \bar{\psi}_j) = \frac{N_c g_{z'}^2 M_{Z^{\prime}}}{24\pi}\left(|B_{ij}^{\psi_L}|^2 + |B_{ij}^{\psi_R}|^2\right) \; ,
\label{zDeWidNew}
\end{equation}
and the ratios of branching fractions for the flavor preserving leptonic decay channels with respective to the $\tau$ decay channels are predicted to be 
\begin{eqnarray} 
Br(Z^{\prime} \rightarrow e^{+}e^{-}) & : & Br(Z^{\prime} \rightarrow \mu^{+} \mu^{-}):Br(Z^{\prime} \rightarrow \tau^{+} \tau^{-})  \\
& \sim & (q_{L_1}^2 + q_{e_1}^2):(q_{L_2}^2 + q_{e_2}^2):(q_{L_3}^2 + q_{e_3}^2) \nonumber \\
& \sim & 37.378:7.153:1 \; . \nonumber
\end{eqnarray}
The branching fractions for different $Z^{\prime}$ hadronic decay channels relative to the $Z^{\prime} \rightarrow \tau^{+} \tau^{-}$ channel are
\begin{eqnarray}
Br(Z^{\prime} \rightarrow u \bar{u}) :  Br(Z^{\prime} \rightarrow d \bar{d}):Br(Z^{\prime} \rightarrow c \bar{c}):Br(Z^{\prime} \rightarrow s \bar{s}):Br(Z^{\prime} \rightarrow t \bar{t}) \qquad \\
\qquad \qquad : Br(Z^{\prime} \rightarrow b \bar{b}):Br(Z^{\prime} \rightarrow \tau^{+} \tau^{-})  \nonumber \\
\sim  3(2|B_{11}^{u_L}|^2):3(q_{Q_1}^2 + q_{d_1}^2):3(2|B_{22}^{u_L}|^2):3(q_{Q_2}^2 + q_{d_2}^2):3(2|B_{33}^{u_L}|^2):3(q_{Q_3}^2 + q_{d_3}^2):(q_{L_3}^2 + q_{e_3}^2)   \nonumber \\
 \sim  31.271:112.134:6.003:21.459:2.595:3:1 \; . \nonumber 
\end{eqnarray}
Due to the large right-handed neutrino charges, the branching fraction for the $Z^{\prime}$ invisible decay is enhanced in our model, 
\begin{equation}
\frac{Br(Z^{\prime} \rightarrow ~\mbox{invisible})}{Br(Z^{\prime} \rightarrow \tau^{+} \tau^{-})}  = 983.62 \; . 
\end{equation}

\subsection{Flavor Violating Decays}

In addition to the differentiable flavor conserving $Z^{\prime}$ decay channels, the $Z^{\prime}$ in our model also has flavor violating decay modes. Specifically, these are the decay modes into different generations of up-type quarks, and the branching fractions of these decays are 
\begin{eqnarray}
Br(Z^{\prime} \rightarrow u \bar{c}, \bar{u} c):Br(Z^{\prime} \rightarrow u \bar{t}, \bar{u} t):Br(Z^{\prime} \rightarrow c \bar{t}, \bar{c} t):Br(Z^{\prime} \rightarrow \tau^{+}\tau^{-}) \\
\sim 1.176:3.061:0.426:1 \; . \nonumber
\end{eqnarray}
The branching fractions of some of the flavor violating $Z^{\prime}$ decay modes are comparable to those of the flavor conserving processes. Thus if  the quark flavors can be identified, we may be able to detect these  flavor violating processes~\cite{ref:Z'FVTop}. 

In addition to the flavor violating $Z^{\prime}$ decay, top quark and charm quark rare decays are also allowed and thus they can potentially distinguish our model from flavor conserving $U(1)^{\prime}$ models. For example, the rare decays $t \rightarrow q l_i \bar{l}_i$ and $t \rightarrow q \nu_i \bar{\nu}_i$ (where $q$ can be any up-type quark except for the top quark) are possible at the tree level through $Z^{\prime}$ exchange  in our model. Using the formulae  given in the appendix, the branching fraction $Br(t \rightarrow u e^{+}e^{-})$ (which is relatively large among the rare decay modes) is roughly $10^{-12} \sim 10^{-13}$, which are too small to be observable. While the branching fractions of the decay mode $t \rightarrow u \nu \bar{\nu}$ can be a factor of $10 \sim 100$ bigger, they are still unobservable at the current experiments. These branching fractions, which scale as $g_{z'}^{4}$, are highly suppressed due to the small overall $U(1)^{\prime}$ gauge coupling $g_{z'}$ for which we choose $g_{z'} = 0.02$ as a benchmark point in this paper. If $g_{z'}$ can be allowed to increase,  while still satisfying all of the electroweak precision measurements and FCNC constraints, by including renormalization group contribution to $Z-Z^{\prime}$ mixing and superpartner contributions to $D^{0}-\overline{D}^{0}$ mixing, these flavor violating decays may be accessible to the current experiments. (Existing phenomenological study of these flavor violating decays can be found in ~\cite{ref:topRareDec}.) 

\section{Sparticle mass spectrum}
\label{sec:susyMass}

Since we extend the MSSM by a generation dependent $U(1)_F^{\prime}$ symmetry, the mass spectrum of the sparticles and their associated phenomenology in our model can be different from the usual MSSM. Below we show the electroweak and $U(1)_F^{\prime}$ symmetry breaking conditions, the msss spectrum of the sparticles, as well as $\beta$ functions of the gauge couplings and the Yukawa couplings, which are modified due to the presence of $U(1)_{F}^{\prime}$. The neutralino sector is the most significantly different sector from that of the MSSM; this is illustrated by a numerical example. 
\subsection{Analytical Result}
\subsubsection{Higgs Sector}
\label{sec:higgs-sector}
The scalar potential giving rise to the masses of the scalar components, ($h_u$, $h_d$, $\phi$, $\phi^{\prime}$), of the $H_u$, $H_d$, $\Phi$, $\Phi^{\prime}$ fields from the superpotential (Eq.(\ref{eqn:sPoten})) is given by,
\begin{eqnarray}
V & = & (M_{H_u}^2 + \mu^2)|h_u|^2 + (M_{H_d}^2 + \mu^2)|h_d|^2 + (M_{\Phi}^2+\mu^{\prime 2})|\phi|^2 + (M_{\Phi^{\prime}}^2+\mu^{\prime 2})|\phi^{\prime}|^2 
\\ 
& & + B \mu (h_u h_d + h.c.) + B^{\prime} \mu^{\prime}(\phi \phi^{\prime} + h.c.) + \frac{1}{8} (g_1^2 + g_2^2)(|h_u|^2 + |h_d|^2) 
\nonumber \\ 
&  & + \frac{1}{2} g_2^2 |h_u h_d|^2 + \frac{1}{2} g_{z'}^2 (q_{H_u}|h_u|^2 + q_{H_d}|h_d|^2 + q_{\Phi}|\phi|^2 + q_{\Phi^{\prime}}|\phi^{\prime}|^2) \; ,
\nonumber
\end{eqnarray}
where the last term is the D term associated with $U(1)_{F}^{\prime}$. 

Minimizing the scalar potential, we obtain,
\begin{eqnarray}
M_{H_u}^2 + |\mu|^2 - B \mu \cot \beta - \frac{1}{2}M_{Z^{\prime}}^2 \cos 2\beta + q_{H_u}C = 0 \, \, , 
\label{eq:min1} \\
M_{H_d}^2 + |\mu|^2 - B \mu \tan \beta + \frac{1}{2}M_{Z^{\prime}}^2 \cos 2\beta + q_{H_d}C = 0 \, \, , 
\label{eq:min2} \\
M_{\Phi}^2 + |\mu^{\prime}|^2 + B^{\prime} \mu^{\prime} \cot \psi + q_{\Phi}C = 0 \, \, , \\
M_{\Phi^{\prime}}^2 + |\mu^{\prime}|^2 + B^{\prime} \mu^{\prime} \tan \psi + + q_{\Phi^{\prime}}C = 0 \, \, ,
\label{eq:min3}
\end{eqnarray}
leading to  
\begin{eqnarray}
|\mu|^2 & = & \frac{1}{2} 
\Big{[}  \frac{ M_{H_u}^2 - M_{H_d}^2 + (q_{H_u} + q_{H_d}) C }{\cos 2\beta}
 - \Big{(} M_{Z}^2 + M_{H_u}^2 + M_{H_d}^2 + (q_{H_u} + q_{H_d}) C \Big{)} \Big{]}  \,, 
\label{eqn:miniNew1} \\
B & = & \frac{1}{2\mu} \Big{[} \Big{(} M_{H_u}^2 + M_{H_d}^2 + 2|\mu|^2 + ( q_{H_u} + q_{H_d} ) 
C \Big{)} \sin 2\beta \Big{]} \, , 
\label{eqn:miniNew2} \\
|\mu^{\prime}|^2 & = & \frac{1}{2} \Big{[} \frac{M_{\Phi}^2 - M_{\Phi^{\prime}}^2 + (q_{\Phi} + q_{\Phi^{\prime}} ) C}{\cos 2\psi}  - \Big{(} M_{\Phi}^2 + M_{\Phi^{\prime}}^2 + ( q_{\Phi} + q_{\Phi^{\prime}} ) C \Big{)} \Big{]} \, ,
\label{eqn:miniNew3} \\
B^{\prime} & = & -\frac{1}{2\mu^{\prime}} \Big{[} \Big{(} M_{\Phi}^2 + M_{\Phi^{\prime}}^2 + 2 |\mu^{\prime}|^2 + (q_{\Phi} + q_{\Phi^{\prime}} \Big{)} C ) \sin 2 \psi \Big{]}  \, , 
\label{eqn:miniNew4} 
\end{eqnarray}
where parameter $C$ is defined in Eq.(\ref{eqn:varA}).

\subsubsection{Neutralino and Chargino sector}

The (Majorana) mass matrix of the neutralinos $(\tilde{B}, \tilde{W}^3, \tilde{H}_d^0, \tilde{H}_u^0, \tilde{B}^{\prime}, \tilde{\Phi}, \tilde{\Phi}^{\prime})$ is given by
\begin{eqnarray}
(\mathcal{M})^{(0)} = \left(\begin{array}{ccccccc} 
\tilde{M}_1 & 0 & -v_d g_1/\sqrt{2} & v_u g_1/\sqrt{2} & 0 & 0 & 0 \\
0 & \tilde{M}_2 & v_d g_2 /\sqrt{2} & -v_u g_2/\sqrt{2} & 0 & 0 & 0 \\ 
-v_d g_1/\sqrt{2} & v_d g_2/\sqrt{2} & 0 & -\mu &\sqrt{2} v_d q_{H_d}g_{z'} & 0 & 0 \\
\sqrt{2} v_u g_1 & -\sqrt{2} v_u g_2 & -\mu & 0 & \sqrt{2} v_u q_{H_u}g_{z'} & 0 & 0 \\
0 & 0 & \sqrt{2} v_d q_{H_d}g_{z'} & \sqrt{2} v_u q_{H_u}g_{z'} & \tilde{M}_1^{\prime} & \sqrt{2} u_{\phi} q_{\phi}g_{z'} & \sqrt{2} u_{\phi} q_{\phi^{\prime}}g_{z'} \\
0 & 0 & 0 & 0 & \sqrt{2} u_{\phi} q_{\phi}g_{z'} & 0 & \mu^{\prime} \\
0 & 0 & 0 & 0 & \sqrt{2} u_{\phi} q_{\phi^{\prime}}g_{z'} & \mu^{\prime} & 0 \end{array} \right) \; , 
\nonumber\\
\end{eqnarray}
in which $\tilde{M}_1$, $\tilde{M}_1^{\prime}$ and $\tilde{M}_2$ are the corresponding gaugino masses for $U(1)_Y$, $U(1)_{F}^{\prime}$ and $SU(2)_L$, respectively. 
The physical neutralino masses $m_{\tilde{N}_i^0} (i = 1-7)$ can be obtained by diagnolizing the mass matrix above, with each physical neutralino given in terms of the composition defined below,
\begin{equation}
\tilde{N}_i = x_i^{\scriptscriptstyle{\tilde{B}}} \tilde{B} + x_i^{\scriptscriptstyle{\tilde{W}^3}} \tilde{W}^3 + x_i^{\scriptscriptstyle{\tilde{H}_d^0}} \tilde{H}_d^0 + x_i^{\scriptscriptstyle{\tilde{H}_u^0}} \tilde{H}_u^0 + x_i^{\scriptscriptstyle{\tilde{B}^{\prime}}} \tilde{B}^{\prime} + x_i^{\scriptscriptstyle{\tilde{\Phi}}} \tilde{\Phi} + x_i^{\scriptscriptstyle{\tilde{\Phi}^{\prime}}} \tilde{\Phi}^{\prime} \; .
\end{equation}
The interactions among neutrilino, femions and sfermions are summarized in Table ~\ref{tbl:interNFSF}.

\begin{table}
\begin{tabular}{c|l} \hline \hline
$\tilde{N}_i \tilde{u}_j^L u_k^R, \, \tilde{N}_i u_j^L \tilde{u}_k^R$ & $x_i^{\scriptscriptstyle{\tilde{H}_u^0}} y_{jk}^u$
\\  \hline
$\tilde{N}_i \tilde{u}_j^L u_k^L$ & $(x_i^{\scriptscriptstyle{\tilde{B}}} q_{Y}^{\scriptscriptstyle{Q_j}} g_1 + \frac{1}{2} x_i^{\scriptscriptstyle{\tilde{W}^3}} g_2 + x_i^{\scriptscriptstyle{\tilde{B}^{\prime}}} q_{Q_j} g_{z'}) \delta_j^k$ 
\\ \hline
$\tilde{N}_i \tilde{u}_j^R u_k^R$ & $(x_i^{\scriptscriptstyle{\tilde{B}}} q_{Y}^{\scriptscriptstyle{u_j}} g_1 + x_i^{\scriptscriptstyle{\tilde{B}^{\prime}}} q_{u_j}g_{z'}) \delta_j^k$
 \\ \hline
$\tilde{N}_i \tilde{d}_j^L d_k^R, \, \tilde{N}_i d_j^L \tilde{d}_k^R$ & $x_i^{\scriptscriptstyle{\tilde{H}_d^0}} y_{jk}^d$ 
\\ \hline
$\tilde{N}_i \tilde{d}_j^L d_k^L$ & $(x_i^{\tilde{B}} q_{Y}^{Q_j} g_1 - \frac{1}{2} x_i^{\tilde{W}^3} g_2 + x_i^{\tilde{B}^{\prime}} q_{Q_j} g_{z'}) \delta_j^k$ \\ \hline
$\tilde{N}_i \tilde{d}_j^R d_k^R$ & $(x_i^{\tilde{B}} q_{Y}^{d_j} g_1 + x_i^{\tilde{B}^{\prime}} q_{d_j} g_{z'}) \delta_j^k$ \\ \hline
$\tilde{N}_i \tilde{\nu}_j^L \nu_k^R, \, \tilde{N}_i \nu_j^L \tilde{\nu}_k^R$ & $x_i^{\scriptscriptstyle{\tilde{H}_u^0}} y_{jk}^{\scriptscriptstyle{\nu}}$ 
\\ \hline
$\tilde{N}_i \tilde{\nu}_j^L \nu_k^L$ & $(x_i^{\tilde{B}} q_{Y}^{L_j} g_1 + \frac{1}{2} x_i^{\scriptscriptstyle{\tilde{W}^3}} g_2 + x_i^{\scriptscriptstyle{\tilde{B}^{\prime}}} q_{L_j} g_{z'}) \delta_j^k$ 
\\ \hline
$\tilde{N}_i \tilde{\nu}_j^R \nu_k^R$ & $x_i^{\scriptscriptstyle{\tilde{B}^{\prime}}} q_{N_j} g_{z'} \delta_j^k$ 
\\ \hline
$\tilde{N}_i \tilde{e}_j^L e_k^R, \, \tilde{N}_i e_j^L \tilde{e}_k^R$ & $x_i^{\scriptscriptstyle{\tilde{H}_d^0}} y_{jk}^{e}$ 
\\ \hline
$\tilde{N}_i \tilde{e}_j^L e_k^L$ & $(x_i^{\tilde{B}} q_{Y}^{L_j} g_1 - \frac{1}{2} x_i^{\scriptscriptstyle{\tilde{W}^3}} g_2 + x_i^{\scriptscriptstyle{\tilde{B}^{\prime}}} q_{L_j} g_{z'}) \delta_j^k$ \\ \hline
$\tilde{N}_i \tilde{e}_j^R e_k^R$ & $(x_i^{\scriptscriptstyle{\tilde{B}}} q_{Y}^{e_j} g_1 + x_i^{\scriptscriptstyle{\tilde{B}^{\prime}}} q_{e_j} g_{z'}) \delta_j^k$ 
\\ \hline \hline
\end{tabular}
\caption{Interactions among neutralinos, fermions and sfermions. The parameters $y_{jk}^{f}$ are the $(i,j)$ entries in the $f$ sector Yukawa matrix, and $q_{Y}^{f}$ are the hypercharge of the field $f$.}
\label{tbl:interNFSF}
\end{table}

In the basis $(\tilde{W}^+, \tilde{H}_u^+)$, $(\tilde{W}^-, \tilde{H}_d^{-})$, the chargino (Dirac) mass matrix is given by
\begin{equation}
\mathcal{M}^{(c)} = 
\left(\begin{array}{cc} 
\tilde{M}_2 & g_2 v_d 
\\ 
g_2 v_u & \mu \end{array}\right). 
\end{equation}

\subsubsection{Sfermion sector}

The mass squared matrices of the sfermions in the $(\tilde{f}_{i\,L}, \tilde{f}_{i\,R})^{T}$ basis are given by,
\begin{eqnarray}
\mathcal{M}_{\tilde{u}}^2 & = & 
\left(
\begin{array}{cc} 
\left(m_{\tilde{Q}}^2 \right)_{ii} + m_{u_i}^2 + (\frac{1}{2} - \frac{2}{3} \sin^2 \theta_w) M_{Z}^2 \cos 2 \beta 
& m_{u_i}\left((A_U)_{ii} - \mu \cot \beta \right)
 \\
m_{u_i}\left((A_U)_{ii} - \mu \cot \beta \right) 
& \left(m_{\tilde{u}}^2\right)_{ii} + m_{u_i}^2 + \frac{2}{3} \sin^2 \theta_w M_{Z}^2 \cos 2\beta 
\end{array} 
\right) 
, 
\end{eqnarray}
\begin{eqnarray}
\mathcal{M}_{\tilde{d}}^2 & = & \left(\begin{array}{cc} \left(m_{\tilde{Q}}^2 \right)_{ii} + m_{d_i}^2 - (\frac{1}{2} - \frac{1}{3} \sin^2 \theta_w)M_{Z}^2 \cos 2 \beta & m_{d_i}\left((A_D)_{ii} - \mu \tan \beta \right) \\
m_{d_i}\left((A_D)_{ii} - \mu \tan \beta \right) 
& \left(m_{\tilde{d}}^2\right)_{ii} + m_{d_i}^2 - \frac{1}{3} \sin^2 \theta_w M_{Z}^2 \cos 2\beta 
 \end{array} \right) , 
\end{eqnarray}
\begin{eqnarray}
\mathcal{M}_{\tilde{l}}^2 & = & \left(\begin{array}{cc} \left(m_{\tilde{L}}^2 \right)_{ii} + m_{e_i}^2 - (\frac{1}{2} - \sin^2 \theta_w)M_{Z}^2 \cos 2 \beta & m_{e_i}\left((A_E)_{ii} - \mu \tan \beta \right) \\
m_{e_i}\left((A_E)_{ii} - \mu \tan \beta \right) 
& \left(m_{\tilde{e}}^2\right)_{ii} + m_{e_i}^2 - \sin^2 \theta_w M_{Z}^2 \cos 2\beta 
\end{array} \right) , 
\end{eqnarray}
\begin{eqnarray}
\mathcal{M}_{\tilde{\nu}}^2 & = & \left(\begin{array}{cc} \left(m_{\tilde{L}}^2 \right)_{ii} + \frac{1}{2}M_{Z}^2 \cos 2 \beta & 0 
 \\
0 & \left(m_{\tilde{\nu}}^2\right)_{ii} 
\end{array} \right) \, .
\end{eqnarray}
In the $6 \times 6$ soft mass matrices shown above, $m_{\tilde{Q}}, \, m_{\tilde{u}}, \, m_{\tilde{d}}, \, m_{\tilde{L}}, \, m_{\tilde{e}}, \, m_{\tilde{\nu}}$ are the soft masses for squarks and sleptons whose explicit forms depend on the specific SUSY breaking mechanism. $m_{u_i}, \, m_{d_i}, \, m_{e_i}, m_{\nu_i}$ are the quark and lepton masses. $A_{U}, \, A_{D}, \, A_{E}$ are the trilinear terms. For $\mathcal{M}_{\tilde{\nu}}^{2}$, we have neglected terms proportional to $m_{\nu}$, which are negligible.  

In general, the effects of the the $U(1)_{F}^{\prime}$ symmetry can manifest in gravity mediated, gauge mediated, and anomaly mediated contributions. Specifically, the soft mass terms can be expressed, schematically in the basis where gluino couplings are diagonal , as
\begin{eqnarray}
\tilde{M}_{LL}^{2} & = & \tilde{m}^{2}_{L} + \tilde{m}^{\prime \; 2}_{L} + x X_{LL} + D_{LL} \; ,
\\
\tilde{M}_{RR}^{2} & = & \tilde{m}^{2}_{R} + \tilde{m}^{\prime \; 2}_{R} + x X_{RR} + D_{RR} \; .
\end{eqnarray}
where $\tilde{M}_{LL}^{2} = m_{\tilde{Q}}^{2}, \; m_{\tilde{L}}^{2}$, and $\tilde{M}_{RR}^{2} = m_{\tilde{u}}^{2}, \; m_{\tilde{d}}^{2}, \; m_{\tilde{e}}^{2}, \; m_{\tilde{\nu}}^{2}$. 
The first terms refer to the gauge mediated SUSY breaking (GMSB) contributions due to SM gauge interactions, which are flavor universal. They can be schematically written as
\begin{equation}
\tilde{m}_{L}^{2} \sim N_{\mbox{\tiny msg}} \sum_{j} \biggl(\frac{\alpha_{j}}{\pi}\biggr)^{2} \biggl(\frac{F}{M_{\mbox{\tiny msg}}}\biggr)^{2} \propto {\bf 1}_{3\times 3} \; ,
\end{equation} 
and similarly for $\tilde{m}_{R}^{2}$, with $N_{\mbox{\tiny msg}}$ being the number of messengers, $M_{\mbox{\tiny msg}}$ being the messenger scale, and $F$ being the F-term of some gauge singlet field that triggers SUSY breaking and the summation is taken over the SM gauge groups under which the matter multiplets are charged. The second terms correspond to the GMSB contributions due to the $U(1)_{F}^{\prime}$ gauge interactions, which are flavor non-universal. These contributions have the form,
\begin{equation}
(\tilde{m}_{L}^{2})_{ii}^{\prime} \sim N_{m} \biggl(\frac{g_{Z^{\prime}}^{2} q_{\psi_{i}}^{2}}{\pi}\biggr)^{2} \biggl(\frac{F}{M_{\mbox{\tiny msg}}}\biggr)^{2} \; ,
\end{equation} 
and similarly for $(\tilde{m}_{R}^{2})_{ii}^{\prime}$, with $q_{\psi_{i}}$ being the $U(1)_{F}^{\prime}$ charges of the corresponding sfermions. The third terms are due to gravity mediation arising from operators in the Kahler potential of the form~\cite{Nir:1993mx},
\begin{equation}
X_{ij} \biggl( \frac{F}{\overline{M}_{\mbox{\tiny Pl}}} \biggr)^{2} L_{i} L_{j}^{\dagger} \; ,
\end{equation}
where $\overline{M}_{\mbox{\tiny Pl}}$ is the reduced Planck scale, and the coefficients $X_{ij}$ are determined by the $U(1)_{F}^{\prime}$ charge differences, $|q_{\psi_{i}} - q_{\psi_{j}}|$. The parameter $x$ characterizes the relative size of the gravity mediated contribution to the gauge mediated contribution. The terms $D_{LL}$ and $D_{RR}$ are the D term contributions, which are $\sim \zeta q_{\psi_{i}}$ where $\zeta$ is a constant. While the D term contributions are flavor diagonal, due to the generation dependent charges, they are flavor non-universal.

For the first two generations of the fermions, since their masses (which are leading terms in the off-diagonal entries in the mass squared matrices that couple $\tilde{f}_{L}$ and $\tilde{f}_{R}$) are very small compared to the soft mass terms (which are the leading terms in the entries that involve $\tilde{f}_{L} \tilde{f}_{L}$ and $\tilde{f}_{R} \tilde{f}_{R}$ in the mass squared matrices), we neglect the left-right mixings.  

As the GMSB contributions associated with $U(1)_{F}^{\prime}$ are non-universal, there can exist CKM and MNS induced FCNCs. Nevertheless, due to the relative smallness of $g_{z^{\prime}}q_{\psi_{i}}$ with respective to SM gauge coupling constants, the flavor non-universal F term contributions are much smaller  compared to the flavor universal F term contributions from GMSB. For the gravity mediated contributions, the $U(1)_{F}^{\prime}$ charges in our model predict that $X_{LL} = \bf{1}_{3\times 3}$ for all sfermions except the up-type squarks ($\tilde{Q}_{i}, \, \tilde{u}_{i}$), and they are, 
\begin{equation}
X_{QQ/uu} = \left(\begin{array}{ccc}
1 & \epsilon^{5/3} & \epsilon^{11/3} \\
\epsilon^{5/3} & 1 & \epsilon^{2} \\
\epsilon^{11/3} & \epsilon^{2} & 1
\end{array}\right) 
\sim 
\left( \begin{array}{ccc}
1 & 0.08 & 0.004 \\
0.08 & 1 & 0.05 \\
0.004 & 0.05 & 1
\end{array}\right)
\; .
\end{equation}
The non-universal D term contributions can also lead to CKM and MNS induced FCNCs. However, due to the relative smallness of $g_{z^{\prime}} q_{\psi_{i}}$, these D terms contributions are much suppressed compared to the SM D term contributions. In anomaly mediated SUSY breaking, the $U(1)_{F}^{\prime}$ effects can manifest through the anomalous dimensions of the matter fields. Due to the smallness of $g_{z^{\prime}} q_{\psi_{i}}$ compared to SM gauge couplings, their contributions to the anomalous dimensions of the matter fields are suppressed.

\subsubsection{Renormalization Group Equations}

The presence of $U(1)_{F}^{\prime}$ also changes the $\beta$ functions. For the beta functions of the gauge coupling constants, the effects of $U(1)_{F}^{\prime}$ appear at the two loop level. For the beta functions of the Yukawa coupling constants, the $U(1)_{F}^{\prime}$ effects appear at one loop. Below are the beta functions including the leading $U(1)_{F}^{\prime}$ effects, 
\begin{eqnarray}
\beta_{g_1^{\prime}} & = & \frac{g_1^{\prime \, 3}}{16 \pi^2} \Big{\{}
\frac{33}{5} + \frac{1}{16 \pi^2} \Big{[} \frac{88}{5} g_3^2 + \frac{27}{5}g_2^2 + \frac{199}{25}g_1^{\prime \, 2} + \frac{12}{5}g_{z'}^2 \mbox{Tr} (q_Y^2q^2) 
\\
& & \hspace{0.5in}
  - \frac{26}{5}Y_{U_3}^2 - \frac{14}{5}Y_{D_3}^2 - \frac{18}{5}Y_{E_3}^2 \Big{]} \Big{\}} \; ,
\nonumber \\
\beta_{g_2} & = & \frac{g_2^3}{16 \pi^2}
\Big{\{} 1 + \frac{1}{16 \pi^2} \Big{[} 24g_3^2 + 25g_2^2 + \frac{9}{5}g_1^{\prime \, 2} + 2g_{z'}^2 (q_{H_u}^2 + q_{H_d}^2 + \sum_i (q_{L_i}^2 + 3q_{Q_i}^2) ) 
\\
& & \hspace{0.5in} - 6Y_{U_3}^2 - 6Y_{D_3}^2 - 2Y_{E_3}^2 \Big{]} \Big{\}} \,\, ,
\nonumber \\
\beta_{g_3} & = & \frac{g_3^3}{16 \pi^2} \Big{\{} -3 + \frac{1}{16\pi^{2}} \Big{[} 14g_3^2 + 9g_2^2 + \frac{11}{5} g_1^{\prime \, 2} + 2g_{z'}^2 \sum_i (2q_{Q_i}^2 + q_{u_i}^2 + q_{d_i}^2) - 4Y_{U_3}^2 - 4Y_{D_3}^2  \Big{]} \Big{ \}}  \,\, ,
\\
\beta_{g_{z'}} & = & \frac{g_{z'}^3}{16 \pi^2} \Big{\{} \mbox{Tr} (q^2) + \frac{1}{16 \pi^2} \Big{[} 16g_3^2 \sum_i (2q_{Q_i}^2 + q_{u_i}^2 + q_{d_i}^2) 
\\
& & \hspace{0.5in} 
+ 6g_2^2 \Big{(} q_{H_u}^2 + q_{H_d}^2 + \sum_i (q_{L_i}^2 + 3q_{Q_i}^2) \Big{)} 
+ \frac{12}{5}g_1^{\prime \, 2} \mbox{Tr} (q_Y^2 q^2) 
+ 4g_{z'}^2 \mbox{Tr} (q^4) 
\nonumber \\
& & \hspace{0.5in}
 - 12(q_{H_u}^2 + q_{Q_3}^2 + q_{u_3}^2)Y_{U_3}^2
 - 12(q_{H_d}^2 + q_{Q_3}^2 + q_{d_3}^2)Y_{D_3}^2 
 - 4(q_{H_d}^2 + q_{L_3}^2 + q_{e_3}^2)Y_{E_3}^2 \Big{]} \Big{\}} \; , 
\nonumber
\end{eqnarray}
where $g_1^{\prime} = \sqrt{\frac{5}{3}} \, g_1$ and $q_Y$ is the $U(1)_Y$ charge of the fermions.

Most of the the $\beta$ functions of the Yukawa couplings except the $(3, 3)$ elements are close to zeros and ignored here. Taking the $U(1)_F^{\prime}$ symmetry into account, the $\beta$ functions of the $(3, 3)$ elements are given by
\begin{eqnarray}
\beta_{Y_{U_3}} & = &
 \frac{Y_{U_3}}{16 \pi^2} \biggl[ 6Y_{U_3}^2 + Y_{D_3}^2 - \frac{16}{3}g_3^2 - 3g_2^2 - \frac{13}{15}g_1^{\prime \, 2} - 2g_{z'}^2(q_{H_u}^2 + q_{Q_3}^2 + q_{u_3}^2) \biggr] \,\, , 
\\
\beta_{Y_{D_3}} & = & \frac{Y_{D_3}}{16 \pi^2} \biggl[ 6Y_{D_3}^2 + Y_{U_3}^2 + Y_{E_3}^2- \frac{16}{3}g_3^2 - 3g_2^2 - \frac{7}{15}g_1^{\prime \, 2} - 2g_{z'}^2(q_{H_d}^2 + q_{Q_3}^2 + q_{d_3}^2) \biggr] \,\, , 
\\
\beta_{Y_{E_3}} & = & \frac{Y_{E_3}}{16 \pi^2} \biggl[ 3Y_{D_3}^2 + 4Y_{E_3}^2- 3g_2^2 - \frac{9}{5}g_1^{\prime \, 2} - 2g_{z'}^2(q_{H_3}^2 + q_{L_3}^2 + q_{e_3}^2) \biggr] \,\, .
\end{eqnarray}

Due to the non-universal $U(1)_{F}^{\prime}$ charges, there exist non-universal contributions to the sfermion soft masses in the RG equations. Nevertheless, the smallness of the $g_{z^{\prime}} q_{\psi_{i}}$ suppresses these non-universal contributions. 

\subsection{Numerical Result}

In the numerical example, we utilize two loop RGEs for all coupling constants.  Given that the effects of $U(1)_{F}^{\prime}$ are subdominant, we restrict ourselves in the numerical example below to the mSUGRA boundary condition for MSSM multiplets, and we include the up type quark mixing. We choose the SPS 1A values for the soft mass parameters at the GUT scale: the universal scalar soft mass $m_0 = 100$ GeV, the universal gaugino soft mass $m_{1/2} = 250$ GeV and the trilinear term $A_{U} = A_{D} = A_{E} = -100$ GeV. In addition, we take $\tan \beta = 25 , \, \tan \psi = 0.9$. Since the gauge coupling of the $U(1)_F^{\prime}$ ($g_{z'}$) is relatively small, it is reasonable to ignore the running effect from the $U(1)_F^{\prime}$. Further, we obtain the soft masses of the gauginos at the SUSY breaking scale which are $\tilde{M}_1 = 101.56$ GeV and $\tilde{M}_2 = 191.8$ GeV. For the soft mass of gaugino $\tilde{B}^{\prime}$, we choose $\tilde{M}_1^{\prime} = 1000$ GeV. The values of $\mu$ and $\mu^{\prime}$ are determined using the minimization conditions given in Eqs. (\ref{eqn:miniNew1})-(\ref{eqn:miniNew4}) and they are $\mu = 904.068$ GeV and $\mu^{\prime} = 1224.7 i$ GeV ($i$ can be rotated away by redefining the scalar fields of the $\Phi$ and $\Phi^{\prime}$).
With these input parameters, the neutralino mass matrix is given by
\begin{eqnarray}
\label{eqn:massNeu}
(\mathcal{M}^{(0)}/\mbox{GeV}) = \left(\begin{array}{ccccccc} 101.56 & 0 & -1.68244 & 42.061 & 0 & 0 & 0 \\
0 & 191.8 & 3.07837 & -76.9594 & 0 & 0 & 0 \\ 
-1.68244 & 3.07837 & 0 & -904.068 & 0.0306316 & 0 & 0 \\
42.061 & -76.9594 & -904.068 & 0 & -8.96168 & 0 & 0 \\
0 & 0 & 0.0306316 & -8.96168 & 1000 & -668.938 & 743.264 \\
0 & 0 & 0 & 0 & -668.938 & 0 & 1224.7i \\
0 & 0 & 0 & 0 & 743.264 & 1224.7i & 0 \end{array} \right)  \; .
\nonumber \\ 
\end{eqnarray}
After diagonalizing the mass matrix above, the masses of the neutralinos and their compositions are summarized in Table~\ref{tbl:massMixNeu}.

\begin{table}[t!]
\begin{tabular}{c|c|c|c|c|c|c|c|c} \hline \hline
Neutralino & Mass (GeV) & $\tilde{B}$ & $\tilde{W}^3$ & $\tilde{H}_d^0$ & $\tilde{H}_u^0$ & $\tilde{B}^{\prime}$ & $\tilde{\Phi}$ & $\tilde{\Phi}^{\prime}$ \\ \hline
$\tilde{N}_1$ & 101.20 & $99.75 \%$ & $0.01 \%$ & $0.23 \%$ & $0.01 \%$ & $0 \%$ & $0 \%$ & $0 \%$ \\ \hline
$\tilde{N}_2$ & 189.82 & $0.01 \%$ & $99.15 \%$ & $0.79 \%$ & $0.05 \%$ & $0 \%$ & $0 \%$ & $0 \%$ \\ \hline
$\tilde{N}_3$ & 907.39 & $0.08 \%$ & $0.23 \%$ & $49.65 \%$ & $50.04 \%$ & $0 \%$ & $0 \%$ & $0 \%$ \\ \hline
$\tilde{N}_4$ & 909.69 & $0.15 \%$ & $0.62 \%$ & $49.33 \%$ & $49.90 \%$ & $0.01 \%$ & $0 \%$ & $0 \%$ \\ \hline
$\tilde{N}_5$ & 1042.36 & $0 \%$ & $0 \%$ & $0 \%$ & $0 \%$ & $24.60 \%$ & $38.04 \%$ & $37.36 \%$ \\ \hline
$\tilde{N}_6$ & 1223.47 & $0 \%$ & $0 \%$ & $0 \%$ & $0 \%$ & $0.07 \%$ & $50.92 \%$ & $49.01 \%$ \\ \hline
$\tilde{N}_7$ & 1515.00 & $0 \%$ & $0 \%$ & $0 \%$ & $0.01 \%$ & $75.38 \%$ & $12.08 \%$ & $12.53 \%$ \\ \hline \hline
\end{tabular}
\caption{Compositions and the mass spectrum of the neutralinos.}
\label{tbl:massMixNeu}
\end{table}

From Table~\ref{tbl:massMixNeu}, we note that the additional neutralinos, $\tilde{N}_{5,6,7}$,  associated with the $U(1)_F^{\prime}$ symmetry are heavier compared to those ($\tilde{N}_{1,2,3,4}$) that exist in the usual MSSM. These additional heavy neutralinos are decoupled from the MSSM. Due to this near block diagonal form of the neutralino mass matrix, the mass spectrum of the light neutralinos $\tilde{N}_{1,2,3,4}$ is very similar to that in the usual MSSM where $\tilde{N}_{1} \simeq \tilde{B}$, $\tilde{N}_{2} \simeq \tilde{W}^{3}$, $\tilde{N}_{3,4} \simeq \frac{1}{2}( \tilde{H}_{u}^{0} \pm \tilde{H}_{d}^{0})$, with $\tilde{N}_{1}$ being the lightest neutralino. 

The mass matrix of the chargino is 
\begin{equation}
(\mathcal{M}^{(c)}/\mbox{GeV}) = 
\left(\begin{array}{cc} 
191.8 & 4.353473 \\ 
108.8370 & 904.068 \end{array}\right). 
\end{equation}

The mass spectrum of the charginos and the sfermions is summarized in Table ~\ref{tbl:massCharSfer}. 
The lightest sfermion in this example is the stau. 

\begin{table}
\begin{tabular}{c|c|c|c|c|c|c|c|c|c|c|c|c|c} \hline \hline
Field & $\tilde{\chi}_1^{\pm}$ & $\tilde{u}_L$ & $\tilde{u}_R$ & $\tilde{c}_L$ & $\tilde{c}_R$ & $\tilde{t}_1$ & $\tilde{t}_2$ & $\tilde{d}_L$ & $\tilde{d}_R$ & $\tilde{s}_L$ & $\tilde{s}_R$ & $\tilde{b}_1$ & $\tilde{b}_2$ \\ \hline
Mass (GeV)& 191.14 & 562.73 & 545.87 & 562.72 & 545.87 & 375.83 & 578.91 & 568.28 & 545.73 & 568.27 & 545.72 & 389.37 & 592.14 \\ \hline \hline
Filed & $\tilde{\chi}_2^{\pm}$ & $\tilde{e}_L$ & $\tilde{e}_R$ & $\tilde{\mu}_L$ & $\tilde{\mu}_R$ & $\tilde{\tau}_1$ & $\tilde{\tau}_2$ & $\tilde{\nu}_{e_L}$ & $\tilde{\nu}_{e_R}$ & $\tilde{\nu}_{\mu_L}$ & $\tilde{\nu}_{\mu_R}$ & $\tilde{\nu}_{\tau_L}$ & $\tilde{\nu}_{\tau_R}$ \\ \hline
Mass (GeV) & 904.73 & 202.56 & 144.20 & 202.56 & 144.16 & 120.68 & 263.61 & 186.10 & 195.53 & 186.08 & 195.51 & 180.33 & 190.46 \\ \hline \hline
\end{tabular}
\caption{The mass pectrum of the charginos and sfermions}
\label{tbl:massCharSfer}
\end{table}

\section{Conclusion}
\label{sec:conclude}
In this paper, we propose a non-universal $U(1)^{\prime}_{F}$ symmetry combined with MSSM.  All gauge anomaly cancellation conditions in our model are satisfied without exotic fields other than three right-handed neutrinos. Because all three generations of chiral superfields have different $U(1)^{\prime}_{F}$ charges,  realistic masses and mixing angles in both the quark and lepton sectors are obtained, after the $U(1)^{\prime}_{F}$ symmetry is broken at a low scale. In our model, neutrinos are predicted to be Dirac fermions and their mass ordering is of the inverted hierarchy type.  The $U(1)^{\prime}_{F}$ charges of the chiral super-fields also naturally suppress the $\mu$ term and automatically forbid baryon number and lepton number violating operators. 
Even though all FCNCs constraints in the down quark and charged lepton sectors can be satisfied, we find that constraint from $D^{0}-\overline{D}^{0}$ turns out to be much more stringent than the constraints from the precision electroweak data.

\begin{acknowledgments}
We thank Daniel Whiteson for useful communication and for providing us the updated CDF limit~\cite{ref:CDFlimit}. The work was supported, in part, by the National Science Foundation under Grant No. PHY-0709742. 
\end{acknowledgments}

\begin{appendix}
\section{Top Quark Rare Decays}
The effective four fermion operators that can lead to top quark rare decays are~\cite{ref:FCNC},
\begin{eqnarray}
  -\mathcal{L}_{eff}&=&\frac{4G_F}{\sqrt{2}}\sum\limits_{m=\psi,\chi}
    \left(\rho_{eff} {J_m}^2 
    + 2wJ_m\cdot J_m^{\prime}+ y {J_m^{\prime}}^2\right)   \label{Leff}  \\[1ex] 
  &=&\frac{4G_F}{\sqrt{2}}\sum_{\psi,\chi}\sum_{i,j,k,l}
  \left[
      C_{kl}^{ij} Q_{kl}^{ij}
    + \tilde{C}_{kl}^{ij} \tilde{Q}_{kl}^{ij}
    + D_{kl}^{ij} O_{kl}^{ij}
    + \tilde{D}_{kl}^{ij} \tilde{O}_{kl}^{ij}
  \right]\;, \nonumber
\end{eqnarray}
where $J_{m}$ and $J_{m}^{\prime}$ are currents that couple to $Z$ and $Z^{\prime}$, respectively, and 
\begin{eqnarray}
  Q_{kl}^{ij}
    =\left(\bar{\psi}_i\gamma^{\mu}P_L\psi_j\right)
    \left(\bar{\chi}_k\gamma_{\mu}P_L\chi_l\right)\;,&\qquad&
  \tilde{Q}_{kl}^{ij}
    =\left(\bar{\psi}_i\gamma^{\mu}P_R\psi_j\right)
    \left(\bar{\chi}_k\gamma_{\mu}P_R\chi_l\right)\;,\label{op}\\[1ex]
  O_{kl}^{ij}
    =\left(\bar{\psi}_i\gamma^{\mu}P_L\psi_j\right)
    \left(\bar{\chi}_k\gamma_{\mu}P_R\chi_l\right)\;,&\qquad&
  \tilde{O}_{kl}^{ij}
    =\left(\bar{\psi}_i\gamma^{\mu}P_R\psi_j\right)
    \left(\bar{\chi}_k\gamma_{\mu}P_L\chi_l\right)\;.\nonumber
\end{eqnarray}
The variables $\psi$ and $\chi$ denote the fermionic fields while $i,j,k,l$ are
the family indices. The coefficients for the effective four fermion operators are
\begin{eqnarray}
  C_{kl}^{ij}&=& \rho_{eff}\delta_{ij}\delta_{kl}\epsilon_i^{\psi_L}\epsilon_k^{\chi_L}
    + w\delta_{ij}\epsilon_i^{\psi_L}B_{kl}^{\chi_L}
    + w\delta_{kl}\epsilon_i^{\chi_L}B_{ij}^{\psi_L}
    + yB_{ij}^{\psi_L}B_{kl}^{\chi_L}\;,\\[1ex]
  \tilde{C}_{kl}^{ij}&=& \rho_{eff}\delta_{ij}\delta_{kl}\epsilon_i^{\psi_R}\epsilon_k^{\chi_R}
    + w\delta_{ij}\epsilon_i^{\psi_R}B_{kl}^{\chi_R}
    + w\delta_{kl}\epsilon_l^{\chi_R}B_{ij}^{\psi_R}
    + yB_{ij}^{\psi_R}B_{kl}^{\chi_R}\;,\\[1ex]
  D_{kl}^{ij}&=& \rho_{eff}\delta_{ij}\delta_{kl}\epsilon_i^{\psi_L}\epsilon_k^{\chi_R}
    + w\delta_{ij}\epsilon_i^{\psi_L}B_{kl}^{\chi_R}
    + w\delta_{kl}\epsilon_l^{\chi_R}B_{ij}^{\psi_L}
    + yB_{ij}^{\psi_L}B_{kl}^{\chi_R}\;,\\[1ex]
  \tilde{D}_{kl}^{ij}&=& \rho_{eff}\delta_{ij}\delta_{kl}\epsilon_i^{\psi_R}\epsilon_k^{\chi_L}
    + w\delta_{ij}\epsilon_i^{\psi_R}B_{kl}^{\chi_L}
    + w\delta_{kl}\epsilon_l^{\chi_L}B_{ij}^{\psi_R}
    + yB^{\psi_R}_{ij}B_{kl}^{\chi_L}\; ,
\end{eqnarray}
where $\rho_{eff}$, $w$, and $y$ are defined as
\begin{eqnarray}
  \rho_{eff}&=&\rho_1\cos^2\theta + \rho_2\sin^2\theta\;,
    \qquad \rho_i={M_W^2\over M_i^2\cos^2\theta_w}\;,\label{rho}\\[1ex]
  w&=&\frac{g_2}{g_1}\sin\theta\cos\theta(\rho_1-\rho_2)\;,\label{w}\\[1ex]
  y&=&\left(\frac{g_2}{g_1}\right)^2(\rho_1 \sin^2 \theta + \rho_2 \cos^2 \theta)\;,\label{y}
\end{eqnarray}
and $M_1 = M_{Z}$ and $M_2 = M_{Z^{\prime}}$.
At the tree level, the decay width of $q_i \rightarrow q_j \psi_k \bar{\psi}_l$ is
\begin{eqnarray}
\Gamma(q_i \rightarrow q_j \psi_k \bar{\psi}_l) = \frac{3N_{c_k}G_F^2 m_{q_i}^5}{48 \pi^3} 
\hspace{4in}
\\
\times \left(|C_{\psi_k \; \psi_l}^{q_j \; q_i} + C_{q_j \; \psi_l}^{\psi_k \; q_i}|^2 + |\tilde{C}_{\psi_k \; \psi_l}^{q_j \; q_i} + \tilde{C}_{q_j \; \psi_l}^{\psi_k \; q_i}|^2 + |D_{\psi_k \; \psi_l}^{q_j \; q_i}|^2 + |D_{q_j \; \psi_l}^{\psi_k \; q_i}|^2 + |\tilde{D}_{\psi_k \; \psi_l}^{q_j \; q_i}|^2 + |\tilde{D}_{q_j \; \psi_l}^{\psi_k \; q_i}|^2 \right) \; .  \nonumber
\end{eqnarray}
If two fermions in the final state are the same ($q_j = \psi_k$), we need to take the permutations into account, which leads to 
\begin{equation}
\Gamma(q_i \rightarrow q_j \psi_k \bar{\psi}_l) = \frac{3N_{c_k}G_F^2 m_{q_i}^5}{48 \pi^3} \left(2|C_{q_j \; \psi_l}^{q_j \; q_i}|^2 + 2|\tilde{C}_{q_j \; \psi_l}^{q_j \; q_i}|^2 + |D_{q_j \; \psi_l}^{q_j \; q_i}|^2 +|\tilde{D}_{q_j \; \psi_l}^{q_j \; q_i}|^2 \right) \; .
\end{equation}
\end{appendix}


\begin{thebibliography}{99}
\bibitem{ref:SO10}
  R. N. Mohapatra, \textit{Unification and Supersymmetry}, Springer, New York, 1986, references therein.
  
\bibitem{ref:E6}
  R. W. Robinett, Phys. Rev. {\bf D26}, 2388 (1982); F. del Aguila, M. Quiros and F. Zwirner, Nucl. Phys. {\bf B284}, 530 (1987); J. L. Hewett and T. G. Rizzo, Phys. Rept. {\bf 183}, 193 (1989).

\bibitem{ref:E6Enrico}  
  E. Nardi, Phys. Rev. {\bf D48}, 3277 (1993); for the phenomenology of the model, see E. Nardi, Phys. Rev. {\bf D49}, 4394 (1994) and E. Nardi, T. G. Rizzo, Phys. Rev. {\bf D50}, 203 (1994).

\bibitem{ref:nonUnifU1}
K. T. Mahanthappa, P. K. Mohapatra, Phys. Rev. {\bf D42}, 2400 (1990).
 
\bibitem{ref:strM}
  C. Coriano, A. E. Faraggi and M. Guzzi, Eur. Phys. J. {\bf C53}, 421 (2008).

\bibitem{ref:PLRevZ'}
 P. Langacker, Rev. Mod. Phys. {\bf 81}, 1199 (2009).
  
\bibitem{ref:Z'LHCRizzo}
  T. G. Rizzo, SLAC-PUB-12129 [arXiv:hep-ph/0610104] TASI 2006 Lecture Notes, and references therein.
  
\bibitem{ref:Z'LHCRizzoInd}
  T. G. Rizzo, JHEP {\bf 08}, 82 (2009).
  
\bibitem{ref:FCNC}
  P. Langacker, M. Plumacher, Phys. Rev. {\bf D62}, 013006 (2000).

\bibitem{ref:frogNiel} 
  C. D. Froggatt, H. B. Nielsen, Nucl. Phys. {\bf B147}, 277 (1979).
  
\bibitem{ref:gauTrmNeuM}
  M.-C. Chen, A. de Gouv\^{e}a, B. A. Dobrescu, 
  Phys. Rev. {\bf D75}, 055009 (2007); M.-C. Chen, J. Huang, 
  Phys.Rev.{\bf D81}, 055007 (2010).
  
\bibitem{ref:SU(5)U(1)} 
  M.-C. Chen, D. R. T. Jones, A. Rajaraman, H. B. Yu, Phys. Rev. {\bf D78}, 015019 (2008).
  
\bibitem{ref:highOpeWalter}  
  F. Bonneta, D. Hernandezb, T. Otac, and W. Winter, JHEP {\bf 10}, 76 (2009).  
  
\bibitem{ref:neutinoPhysics}
 Thomas Schwetz, M. A. Tortola, Jose W. F. Valle, New J. Phys. {\bf 10} 113011 (2008).

\bibitem{ref:neutrinoTheta13}
 M.C. Gonzalez-Garcia, M. Maltoni, J. Salvado, JHEP 1004:056 (2010).
  
\bibitem{ref:muParaDawson}
 For reviews, see, for example,  S. Dawson, hep-ph/9712464; 
  S. P. Martin, hep-ph/9709356v5.
  
\bibitem{ref:muProblem}
 G. Cleaver, M. Cvetic, J. R. Espinosa, L. Everett, P. Langacker, Phys. Rev. {\bf D57}, 2701 (1998).  
  
\bibitem{ref:susyPrimer}
S. P. Martin, hep-ph/9709356

\bibitem{ref:ewCons}
  J. Erler, P. Langacker, Phys. Lett. {\bf B667}, 1 (2008). 
  
\bibitem{ref:mixZZ}
  For a recent review, see, G. Altarelli and M. W. Grunewald, Phys. Rept. {\bf 403-404}, 189 (2004) (hep-ph/0404165); DELPHI Collaboration, P. Abreu \textit{et al.}, Z. Phys. C {\bf 65}, 163 (1995); ALEPH, DELPHI, L3, OPAL and SLD Collaborations; LEP EW Working Group; and SLD EW and Heavy Flavour Groups, Phys. Rept. {\bf 427}, 257 (2006). 

\bibitem{ref:CDFz'Con1}
  M. Carena, A. Daleo, B. A. Dobrescu, T. M. P. Tait, Phys. Rev. {\bf D70}, 093009 (2004).
  
\bibitem{ref:rgeLoop}
 K. S. Babu, C. Kolda, J. March-Russell, Phys. Rev. {\bf D54}, 4635(1996).

\bibitem{ref:flavorPhysics}
 A. J. Buras, hep-ph/0910.1032.
  
\bibitem{ref:pdgDDbar}
 Particle Data Group (C. Amsler et al.), Phys.Lett. {\bf B667}, 1 (2008).  
 
\bibitem{ref:FCNCSUSY}
 F. Gabbiani, E. Gabrielli, A. Masiero, L. Silvestrini, Nucl. Phys. {\bf B477}, 321-352 (1996). 
 
\bibitem{ref:Z'FVTop}
Abdesslam Arhrib, Kingman Cheung, Cheng-Wei Chiang, Tzu-Chiang Yuan,  Phys. Rev. {\bf D73},  075015 (2006) and references therein.

\bibitem{ref:CDFlimit}
http://www-cdf.fnal.gov/~danielw/zprime/plots/index.html

\bibitem{ref:topRareDec}
 O. Cakir, I.T. Cakir, A. Senol, T. Tasci, hep-ph/1003.3156.  

\bibitem{Nir:1993mx}
  Y.~Nir and N.~Seiberg,
  Phys.\ Lett.\  B {\bf 309}, 337 (1993);
  J.~L.~Feng, C.~G.~Lester, Y.~Nir and Y.~Shadmi,
  Phys.\ Rev.\  {\bf D77}, 076002 (2008).


\end{thebibliography}
\end{document}